\newcommand{\DoPrePrint}{1} % 0 for 2-column submission/review format; 1 for double-spaced, line-numbered preprint
\ifnum\DoPrePrint=1
  \RequirePackage{lineno} 
  \documentclass[aps,prl,preprint,showpacs,superscriptaddress,nofootinbib,linenumbers]{revtex4}

  \usepackage{lineno}
\else
  \documentclass[aps,prl,twocolumn,showpacs,superscriptaddress,nofootinbib]{revtex4}
\fi

\usepackage{amsmath}   % need for subequations
\usepackage{graphicx}  % for figures
\usepackage{xspace}
\usepackage{units}

\usepackage{hyperref}
\usepackage{multirow}
\usepackage{gensymb}

\newcommand{\minerva}{MINERvA\xspace}
\newcommand{\minos}{MINOS\xspace}

\newcommand{\genie}{GENIE\xspace}
\newcommand{\nuwro}{NuWro\xspace}

\newcommand{\numu}{\ensuremath{\nu_{\mu}}\xspace}
\newcommand{\numubar}{\ensuremath{\bar{\nu}_{\mu}}\xspace}

\newcommand{\QSqproton}{\ensuremath{Q^{2}_{p}}\xspace}

\newcommand{\xsecproton}{\ensuremath{d\sigma/dQ^{2}_{p}}\xspace}

\newcommand{\sizecheck}{0} % 0 to do nothing; 1 to check size
\newcommand{\PRLsupp}{1}   % 0 to do nothing; 1 to put the appendix in a supplement

\ifnum\PRLsupp=0
  
\else
  
\fi

\newif\ifpdf
\ifx\pdfoutput\undefined
   \pdffalse
\else
   \pdfoutput=1
   \pdftrue
\fi
\ifpdf
   \usepackage{graphicx}
   \usepackage{epstopdf}
\else
   \usepackage{graphicx}
\fi

\begin{document}

\ifnum\DoPrePrint=1
%\linenumbers
\fi

\title{Direct Measurement of Nuclear Dependence of Charged Current Quasielastic-like Neutrino Interactions using MINERvA}
%% THIS IS THE AUTHOR LIST FROM THE CCQE ANTI-NEUTRINO PAPER

%% MANUAL PARTS OF AUTHOR LIST

%% (1) need to add ``\thanks{\deceased}'' after DeMaat, Gobbi, Tzanakos
\newcommand{\deceased}{Deceased}

%% (2) we offered Jan authorship, so here it is
%% have to put his author line in by hand
\newcommand{\wroclaw}{Institute of Theoretical Physics, Wroc\l aw University, Wroc\l aw, Poland}    
%% \author{J.T.~Sobczyk}                     \affiliation{\wroclaw}  \affiliation{\FNAL}

%% (2b) Also technical authors non in Glaucus who did not sign the communications paper.  Basically the W&M crowd are the only ones who have requested this

%% AUTOMATIC LIST (EDITED AS ABOVE)
%% List of institution addresses, in command form.
\newcommand{\Rutgers}{Rutgers, The State University of New Jersey, Piscataway, New Jersey 08854, USA}
\newcommand{\Hampton}{Hampton University, Dept. of Physics, Hampton, VA 23668, USA}
\newcommand{\Dortmund}{Institute of Physics, Dortmund University, 44221, Germany }
\newcommand{\Otterbein}{Department of Physics, Otterbein University, 1 South Grove Street, Westerville, OH, 43081 USA}
\newcommand{\JMU}{James Madison University, Harrisonburg, Virginia 22807, USA}
\newcommand{\Florida}{University of Florida, Department of Physics, Gainesville, FL 32611}
\newcommand{\UCIrvine}{Department of Physics and Astronomy, University of California, Irvine, Irvine, California 92697-4575, USA}
\newcommand{\CBPF}{Centro Brasileiro de Pesquisas F\'{i}sicas, Rua Dr. Xavier Sigaud 150, Urca, Rio de Janeiro, Rio de Janeiro, 22290-180, Brazil}
\newcommand{\PUCP}{Secci\'{o}n F\'{i}sica, Departamento de Ciencias, Pontificia Universidad Cat\'{o}lica del Per\'{u}, Apartado 1761, Lima, Per\'{u}}
\newcommand{\INRM}{Institute for Nuclear Research of the Russian Academy of Sciences, 117312 Moscow, Russia}
\newcommand{\Jlab}{Jefferson Lab, 12000 Jefferson Avenue, Newport News, VA 23606, USA}
\newcommand{\Pittsburgh}{Department of Physics and Astronomy, University of Pittsburgh, Pittsburgh, Pennsylvania 15260, USA}
\newcommand{\Guanajuato}{Campus Le\'{o}n y Campus Guanajuato, Universidad de Guanajuato, Lascurain de Retana No. 5, Colonia Centro, Guanajuato 36000, Guanajuato M\'{e}xico.}
\newcommand{\Athens}{Department of Physics, University of Athens, GR-15771 Athens, Greece}
\newcommand{\Tufts}{Physics Department, Tufts University, Medford, Massachusetts 02155, USA}
\newcommand{\WM}{Department of Physics, College of William \& Mary, Williamsburg, Virginia 23187, USA}
\newcommand{\FNAL}{Fermi National Accelerator Laboratory, Batavia, Illinois 60510, USA}
\newcommand{\Purdue}{Department of Chemistry and Physics, Purdue University Calumet, Hammond, Indiana 46323, USA}
\newcommand{\MCLA}{Massachusetts College of Liberal Arts, 375 Church Street, North Adams, MA 01247}
\newcommand{\UMD}{Department of Physics, University of Minnesota -- Duluth, Duluth, Minnesota 55812, USA}
\newcommand{\Northwestern}{Northwestern University, Evanston, Illinois 60208}
\newcommand{\UNI}{Universidad Nacional de Ingenier\'{i}a, Apartado 31139, Lima, Per\'{u}}
\newcommand{\Rochester}{University of Rochester, Rochester, New York 14627 USA}
\newcommand{\Austin}{Department of Physics, University of Texas, 1 University Station, Austin, Texas 78712, USA}
\newcommand{\USM}{Departamento de F\'{i}sica, Universidad T\'{e}cnica Federico Santa Mar\'{i}a, Avenida Espa\~{n}a 1680 Casilla 110-V, Valpara\'{i}so, Chile}
\newcommand{\Geneva}{University of Geneva, 1211 Geneva 4, Switzerland}
\newcommand{\Chicago}{Enrico Fermi Institute, University of Chicago, Chicago, IL 60637 USA}
\newcommand{\hired}{}
\newcommand{\OregonState}{Department of Physics, Oregon State University, Corvallis, Oregon 97331, USA}
\newcommand{\oxford}{}
\newcommand{\umiss}{University of Mississippi, Oxford, Mississippi 38677, USA}
\newcommand{\upenn}{209 S. 33rd St. Philadelphia, PA 19104}
\newcommand{\AMU}{AMU Campus, Aligarh, Uttar Pradesh 202001, India}
\newcommand{\damartinThanks}{now at Illinois Institute of Technology, Chicago, IL 60616, USA}
\newcommand{\higueraThanks}{now at University of Houston, Houston, TX 77204, USA}
\newcommand{\chrismarshallThanks}{now at Lawrence Berkeley National Laboratory, Berkeley, CA 94720, USA}
\newcommand{\mcgivernThanks}{now at Fermi National Accelerator Laboratory, Batavia, IL 60510, USA}
\newcommand{\joelmousseauThanks}{now at University of Michigan, Ann Arbor, MI 48109, USA}
\newcommand{\twaltonThanks}{now at Fermi National Accelerator Laboratory, Batavia, IL 60510, USA}
\newcommand{\jwolcottThanks}{now at Tufts University, Medford, MA 02155, USA}
\newcommand{\janThanks}{also at Institute of Theoretical Physics, Wroc\l aw University, Wroc\l aw, Poland}

% 89 total signatories.
\author{M.~Betancourt}        \affiliation{\FNAL}
\author{A.~Ghosh}            \affiliation{\USM}
\author{T.~Walton}            \affiliation{\FNAL}

\author{O.~Altinok}                       \affiliation{\Tufts}
\author{L.~Bellantoni}                    \affiliation{\FNAL}
\author{A.~Bercellie}                     \affiliation{\Rochester}
\author{A.~Bodek}                         \affiliation{\Rochester}
\author{A.~Bravar}                        \affiliation{\Geneva}
\author{T.~Cai}                           \affiliation{\Rochester}
\author{D.A.~Martinez~Caicedo}\thanks{\damartinThanks}  \affiliation{\CBPF}  \affiliation{\FNAL}
\author{M.F.~Carneiro}                    \affiliation{\OregonState}
\author{S.A.~Dytman}                      \affiliation{\Pittsburgh}
\author{G.A.~D\'{i}az~}                   \affiliation{\Rochester}  \affiliation{\PUCP}
\author{J.~Felix}                         \affiliation{\Guanajuato}
\author{L.~Fields}                        \affiliation{\FNAL}  \affiliation{\Northwestern}
\author{R.~Fine}                          \affiliation{\Rochester}
\author{R.Galindo}                        \affiliation{\USM}
\author{H.~Gallagher}                     \affiliation{\Tufts}
\author{A.~Ghosh}                         \affiliation{\USM}  \affiliation{\CBPF}
\author{T.~Golan}                         \affiliation{\wroclaw}  \affiliation{\Rochester}
\author{R.~Gran}                          \affiliation{\UMD}
\author{D.A.~Harris}                      \affiliation{\FNAL}
\author{A.~Higuera}\thanks{\higueraThanks}  \affiliation{\Rochester}  \affiliation{\Guanajuato}
\author{K.~Hurtado}                       \affiliation{\CBPF}  \affiliation{\UNI}
\author{M.~Kiveni}                        \affiliation{\FNAL}
\author{J.~Kleykamp}                      \affiliation{\Rochester}
\author{T.~Le}                            \affiliation{\Tufts}  \affiliation{\Rutgers}
\author{E.~Maher}                         \affiliation{\MCLA}
\author{S.~Manly}                         \affiliation{\Rochester}
\author{W.A.~Mann}                        \affiliation{\Tufts}
\author{C.M.~Marshall}\thanks{\chrismarshallThanks}  \affiliation{\Rochester}
\author{K.S.~McFarland}                   \affiliation{\Rochester}  \affiliation{\FNAL}
\author{C.L.~McGivern}\thanks{\mcgivernThanks}  \affiliation{\FNAL}  \affiliation{\Pittsburgh}
\author{A.M.~McGowan}                     \affiliation{\Rochester}
\author{B.~Messerly}                      \affiliation{\Pittsburgh}
\author{J.~Miller}                        \affiliation{\USM}
\author{J.G.~Morf\'{i}n}                  \affiliation{\FNAL}
\author{J.~Mousseau}\thanks{\joelmousseauThanks}  \affiliation{\Florida}
\author{D.~Naples}                        \affiliation{\Pittsburgh}
\author{J.K.~Nelson}                      \affiliation{\WM}
\author{A.~Norrick}                       \affiliation{\WM}
\author{Nuruzzaman}                       \affiliation{\Rutgers}  \affiliation{\USM}
\author{C.E.~Patrick}                     \affiliation{\Northwestern}
\author{G.N.~Perdue}                      \affiliation{\FNAL}  \affiliation{\Rochester}
\author{M.A.~Ram\'{i}rez}                 \affiliation{\Guanajuato}
\author{L.~Ren}                           \affiliation{\Pittsburgh}
\author{D.~Rimal}                         \affiliation{\Florida}
\author{P.A.~Rodrigues}                   \affiliation{\umiss}  \affiliation{\Rochester}
\author{D.~Ruterbories}                   \affiliation{\Rochester}
\author{H.~Schellman}                     \affiliation{\OregonState}  \affiliation{\Northwestern}
\author{J.T.~Sobczyk}                     \affiliation{\wroclaw}
\author{C.J.~Solano~Salinas}              \affiliation{\UNI}
\author{S.~S\'{a}nchez~Falero}            \affiliation{\PUCP}
\author{E.~Valencia}                      \affiliation{\WM}  \affiliation{\Guanajuato}
\author{T.~Walton}\thanks{\twaltonThanks}  \affiliation{\Hampton}
\author{J.~Wolcott}\thanks{\jwolcottThanks}  \affiliation{\Rochester}
\author{M.Wospakrik}                      \affiliation{\Florida}
\author{B.~Yaeggy}                        \affiliation{\USM}

%% END AUTOMATIC PART
\collaboration{The \minerva  Collaboration}\ \noaffiliation

\date{\today}

%nubar CCQE paper \pacs{13.15.+g,25.30.Pt,21.10.-k}

\begin{abstract}
Charged-current $\nu_{\mu}$ interactions on carbon, iron, and lead with a final state hadronic system of one or more protons with zero mesons are used to investigate the influence of the nuclear environment on quasielastic-like interactions.    The transfered four-momentum squared to the target nucleus, $Q^2$,  is reconstructed based on the kinematics of the leading proton, and differential cross sections versus $Q^2$ and the cross-section ratios of iron, lead and carbon to scintillator are measured for the first time in a single experiment. The measurements show a dependence on atomic number.  While the quasielastic-like scattering on carbon is compatible with predictions, the trends exhibited by scattering on iron and lead favor a prediction with intranuclear rescattering of hadrons accounted for by a conventional particle cascade treatment. These measurements help discriminate between different models of both initial state nucleons and final state interactions used in the neutrino oscillation experiments.

\end{abstract}

\ifnum\sizecheck=0  
\maketitle
\fi

%%\section{Introduction}

Accurate neutrino cross-section measurements and modeling of nuclear effects are required for precise measurements of neutrino oscillation physics such as CP-violation and the ordering of the neutrino masses~\cite{Abe:2011ks,Ayres:2004js,Acciarri:2015uup}. 
One of the most important channels for lower neutrino energy oscillation experiments is charged current quasielastic (CCQE) scattering, in which a neutrino exchanges a $W$ with a neutron, producing one lepton and one proton in the final state, along with a possibly excited nucleus which typically is undetected. Experiments such as T2K and MiniBooNE use CCQE interactions as the main channel for oscillation measurements ~\cite{Abe:2015ibe,Aguilar-Arevalo:2013pmq} because in principle the neutrino energy can be deduced using only the lepton kinematics and assuming a 2-body elastic scatter. Better understanding of the CCQE process as it occurs in a nuclear medium is also needed for the higher neutrino energies of NO$\nu$A~\cite{Adamson:2016tbq} and the future long baseline oscillation experiments DUNE and HyperK~\cite{Acciarri:2015uup,Ayres:2004js}.

Experiments used large, deuterium-filled bubble chambers in the 1970s--1990s to measure CCQE scattering on (quasi-) free nucleons and obtained consistent results \cite{Baker:1981su,Miller:1982qi,PhysRevD.28.436}. Recent experiments using heavier nuclei such as carbon, oxygen, and iron as targets have shown that our understanding of CCQE is incomplete~\cite{Gran:2006jn,AguilarArevalo:2008rc,AguilarArevalo:2010wv,PhysRevLett.111.022502,Fields:2013zhk,PhysRevD.91.012005}. Using heavier nuclei requires consideration of the role that the nuclear environment plays; the initial state neutrons are in a bound system, and the reaction products can interact on their way out of the nucleus.  A variety of initial state effects have been suggested, including Fermi motion~\cite{Smith:1972xh}, effects resulting in two particles and two holes (2p2h) such as meson exchange currents ~\cite{Martini:2009uj,Nieves:2013fr,PhysRevC.83.045501}, short range correlations ~\cite{Benhar:2006wy,Arrington:2011xs,Arrington:2012ax}, and long range correlations as estimated using the Random Phase Approximation (RPA) ~\cite{Graczyk:2003ru,Nieves:2004wx,Valverde:2006yi}. Final state interactions (FSI), where a hadron scatters on its way out of the nucleus, are modeled in a variety of implementations~\cite{Andreopoulos:2009rq,Nowak:2006xv,Golan:2013jtj}.

Nuclear effects modify both the final-state particle kinematics and content, altering the rate of detection of any given interaction channel.
The neutron's initial state affects the final-state particle kinematics, while FSI affect both kinematics and the particle content of the final state, since particles can be rescattered or absorbed in the nucleus. As a result, a sample of events with an observed lepton and nucleons in the final state may have originated via inelastic processes at production. For example,  $\Delta$ (1232)  resonance  production  and  decay,  if the pion is absorbed during FSI, leaves a QE-like final state that contains only one lepton and some number of nucleons, but no pions. In this case the neutrino energy reconstruction assuming simple two-body kinematics will be incorrect.  It is important that these $A$-dependent nuclear effects, which impact neutrino energy reconstruction as well as signal efficiencies, be understood and modeled since future oscillation experiments will use targets that range from carbon~\cite{Adamson:2016tbq} to argon~\cite{Acciarri:2015uup}.

 Previous measurements from \minerva used the kinematics of the muon or the proton to study quasielastic interactions on CH \cite{PhysRevLett.111.022502,Walton:2014esl}. Protons are affected by FSI, so measuring them provides new constraints on the FSI models. A direct way to elicit nuclear effects in neutrino interactions is by making simultaneous measurements on different nuclei in the same detector. This approach allows flux and detector uncertainties to be reduced by taking ratios of measurements on different nuclei. The first measurement of this kind to be based entirely upon CCQE-like interactions is presented here.

%%\section{\minerva Experiment}

 \minerva uses the neutrinos from the NuMI beamline at Fermi National Accelerator Laboratory~\cite{Adamson:2015dkw}. 
The neutrino flux for the data presented here is peaked at 3 GeV and contains $95\%$ \numu, with the remainder consisting of \numubar, $\nu_e$, and $\bar\nu_e$ ~\cite{Alt:2006fr}. The neutrino beam is simulated with Geant4 9.2.p03~\cite{Agostinelli:2002hh}, and constrained with thin-target hadron production measurements and an in-situ neutrino electron scattering constraint~\cite{Aliaga:2016oaz}. The analysis uses data collected between March 2010 and April 2012, and corresponds to $3.06\times{10}^{20}$ protons on target (POT).

The MINER$\nu$A detector~\cite{Aliaga:2013uqz} is segmented longitudinally into several regions: nuclear targets, the scintillator tracker, and downstream electromagnetic and hadronic calorimeters. The nuclear target region contains five solid passive targets of carbon(C), iron(Fe), and lead(Pb), separated from each other by 4 or 8 scintillator planes for vertex and particle reconstruction. Targets 1, 2 and 3 contain distinct segments of Fe and Pb planes that are 2.6 cm thick; target 3 also has a C segment which is 7.6 cm thick and target 5 has Fe and Pb segments which are 1.3 cm thick. The analysis restricts to targets with 8 scintillator planes on both sides, except target 5 which has 4 scintillator planes upstream the target. The tracker is made solely of scintillator planes. Strips in adjacent planes are rotated by $60^{\circ}$ from each other, which enables three-dimensional track reconstruction~\cite{Aliaga:2013uqz}. The MINOS Near Detector is two meters downstream of the MINER$\nu$A detector and serves as a magnetized muon spectrometer~\cite{Michael:2008bc}.

The neutrino event generator \genie 2.8.4~\cite{Andreopoulos:2009rq} is used to simulate neutrino interactions in the detector. The CCQE scattering model uses a relativistic Fermi gas model (RFG) and dipole axial form factor with $M_A=0.99$ GeV. Resonant production is modeled using the Rein-Sehgal model~\cite{Rein:1980wg}, deep inelastic scattering (DIS) kinematics is modeled using 2003 Bodek-Yang model~\cite{Bodek:2002ps} and the hadron final states are modeled with Koba-Nielsen-Olsen scaling and PYTHIA~\cite{koba,pythia}. The default GENIE simulation has been augmented to include interactions resulting in two particles and two holes (2p2h), as formulated in the Valencia model \cite{Nieves:2011yp,Gran:2013kda,Schwehr:2016pvn}. The relative strength of this 2p2h prediction has been tuned to \minerva inclusive scattering data~\cite{Rodrigues:2015hik}. The RPA effect from the calculation of~\cite{Nieves:2004wx} is included for quasielastic events. Moreover, the GENIE nonresonant pion production prediction has been modified to agree with deuterium data~\cite{Wilkinson:2014yfa}. To treat FSI, \genie uses an effective model, in which hadronic intranuclear rescattering cross sections increase with the nuclear size according to $A^{2/3}$ scaling~\cite{Dytman:2011zz}. The final state distributions in energy and angle come from 2-body kinematics and phase space formulations.  The model provides a good description of hadron-nucleus data. Comparisons are also made to predictions of NuWro event generator~\cite{Golan:2012wx}, which uses a local Fermi gas model, an intranuclear cascade of hadronic interactions in the FSI model and medium corrections~\cite{Salcedo:1987md}. Coulomb corrections are not included in the simulations~\cite{Coulomb}.

 The interactions and decays of particles produced in the neutrino interactions of the final-state particles that exit the nucleus are simulated by Geant4 9.4.2 ~\cite{Agostinelli:2002hh}. The visible energy scale is calibrated using through-going muons, such that the energy deposit (per plane) distribution is the same for data and simulation.  Measurements made with a smaller version of the \minerva detector in a
hadron test beam~\cite{Aliaga:2013uqz} are used to constrain the
uncertainties associated with the detector responses to both protons and charged pions.

%\section{Event Reconstruction and Analysis} 
For this measurement, the QE-like signal is defined as an event with one muon, no pions and at least one proton with momentum greater than 450 MeV/c exiting the nucleus. A sample of QE-like interactions is selected with at least two reconstructed tracks, one from a muon candidate and at least one proton candidate that stops in the detector. The sample includes both muon tracks that exit the sides of the \minerva detector and those that are matched to a track in the \minos detector. The analysis requires events with a reconstructed interaction vertex in the C, Fe or Pb targets or in the fiducial volume of scintillator.  The event selection uses $\mathrm{d}E/\mathrm{d}x$ to identify protons and to estimate their momentum. The $\mathrm{d}E/\mathrm{d}x$ profile of each hadron is fit to templates for pion and proton hypotheses, and the $\chi^2$ used to determine the particle ID~\cite{Twalton-thesis}.

 Backgrounds from inelastic interactions that produce an untracked pion are reduced by cutting on extra energy $E_{extra}$ that is not linked to a track and is located outside of a 10 cm radius sphere centered at the vertex. The cut on $E_{extra}$ is a function of $Q^2_p$, and is described below. This region around the vertex is excluded to avoid bias due to the mismodeling of low energy nucleons near the vertex~\cite{PhysRevLett.111.022502}. Pions of low kinetic energy are removed by a cut on events with Michel electrons from pion decays near the vertex. A similar event selection criteria have been used in a previous publication for events with interactions in the scintillator \cite{Walton:2014esl}. The present measurement uses the experiment$^{'}$s most current flux prediction~\cite{Aliaga:2013uqz}.

From the measured energy and direction of the proton and muon, the four-momentum transfer \QSqproton and angle between the $\nu$-$\mu$ and $\nu$-p reaction planes, coplanarity angle $\varphi$, are reconstructed.
%To compute the $\QSqproton$ we use the leading proton and assume 
In the case of CCQE scattering from a neutron at rest, \QSqproton can be calculated using the proton kinetic energy, $T_p$ alone.  Under this assumption
%%%%%%%%%%%%%%%%%%%%%%%%%%%%%%%%%%%%%%%%%%%%%%%%%%%%
\begin{equation*}
\QSqproton=(M_{n}-\epsilon_B)^2-M^2_{p}+2(M_{n}-\epsilon_{B})(T_{p}+M_{p}-M_{n}+\epsilon_{B}),
\end{equation*}
%%%%%%%%%%%%%%%%%%%%%%%%%%%%%%%%%%%%%%%%%%%%%%%%%%%%%
where $M_{n,p}$ is the nucleon mass, and $\epsilon_B$ is the effective binding energy of +34 MeV/c$^2$ taken up by liberated nucleons~\cite{Moniz:1971mt}. 

\begin{figure}[htpb]
\centering
\includegraphics[trim =  0mm 6mm 15mm 2mm, clip, width=0.48\columnwidth]{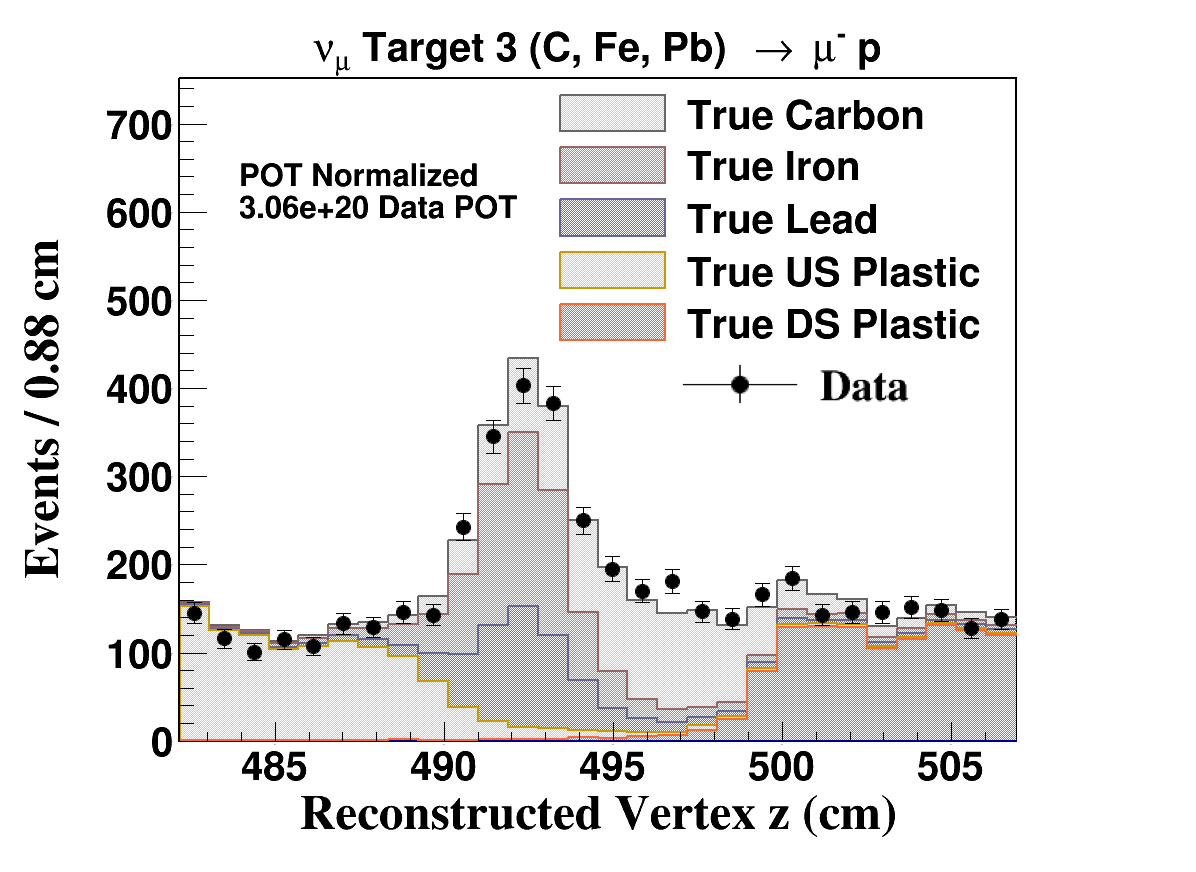}
\includegraphics[trim =  0mm 6mm 15mm 2mm, clip, width=0.48\columnwidth]{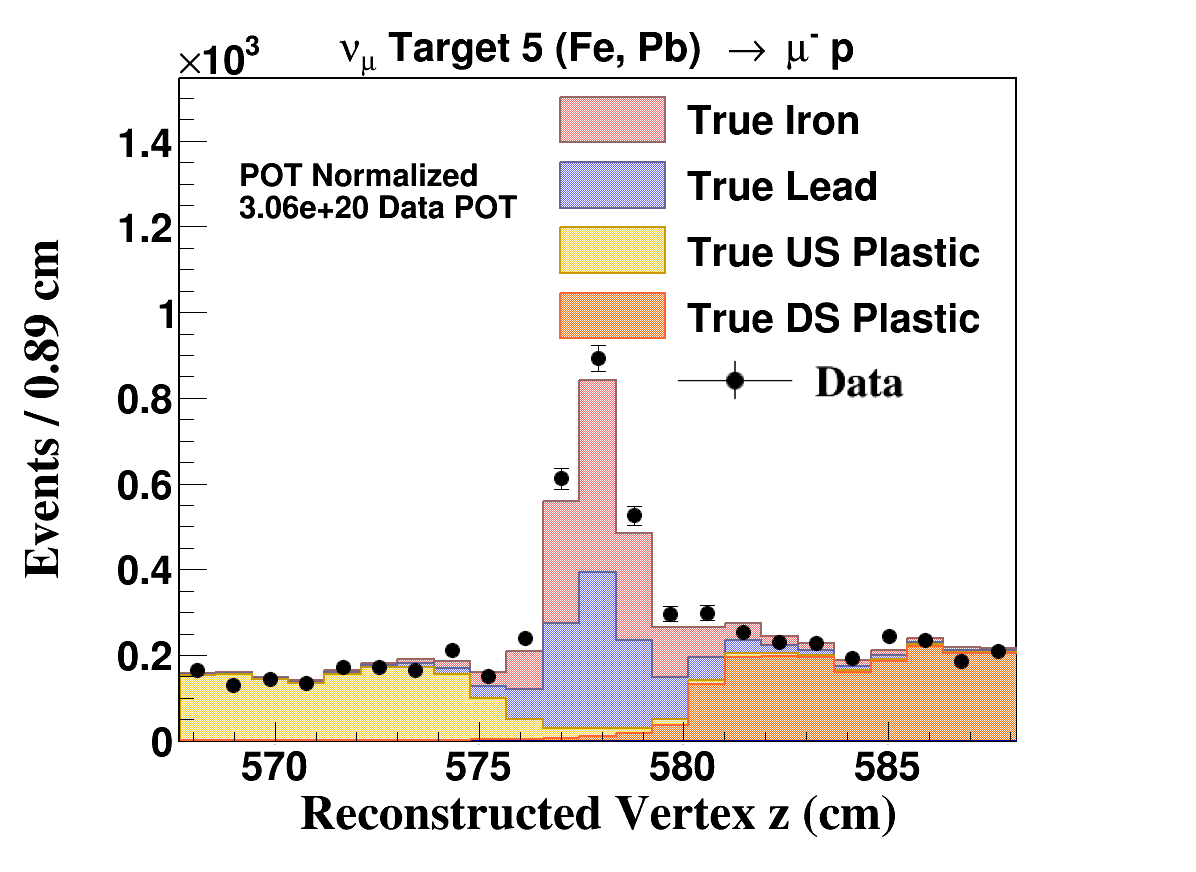}
\caption{Reconstructed interaction vertex $z$ for Target 3 (left) and Target 5 (right). Data and predictions after background tuning are shown for events in each of nucleus. Also shown are contributions from events in the upstream and downstream plastic scintillator.}
\label{targets}
\end{figure}

The selected events in the nuclear targets contain two backgrounds, both of which are constrained using data. The first background consists of interactions incorrectly reconstructed in the nuclear targets that originate in the scintillator surrounding the targets.  Figure \ref{targets} shows the simulated scintillator background, as well as the signal from two different nuclear targets (Target 3 and Target 5) as a function of the reconstructed vertex position along the detector longitudinal axis. Scintillator background events are less than $8\%$ of the final sample. 
The level of this background has been constrained by fitting the tails of the vertex distributions in the upstream (US Plastic) and downstream (DS Plastic) regions for each target subsection separately and extracting a scale factor for the scintillator background. Background scale factors span the range from $0.95 \pm 0.05$ (Target 5) to $1.10 \pm 0.05$ (Target 2). The scale factors are applied to the simulated prediction of the scintillator background in the selected sample.

\begin{figure}[htpb]
\centering
\includegraphics[trim = 0mm 6mm 15mm 2mm, clip, width=0.46\columnwidth]{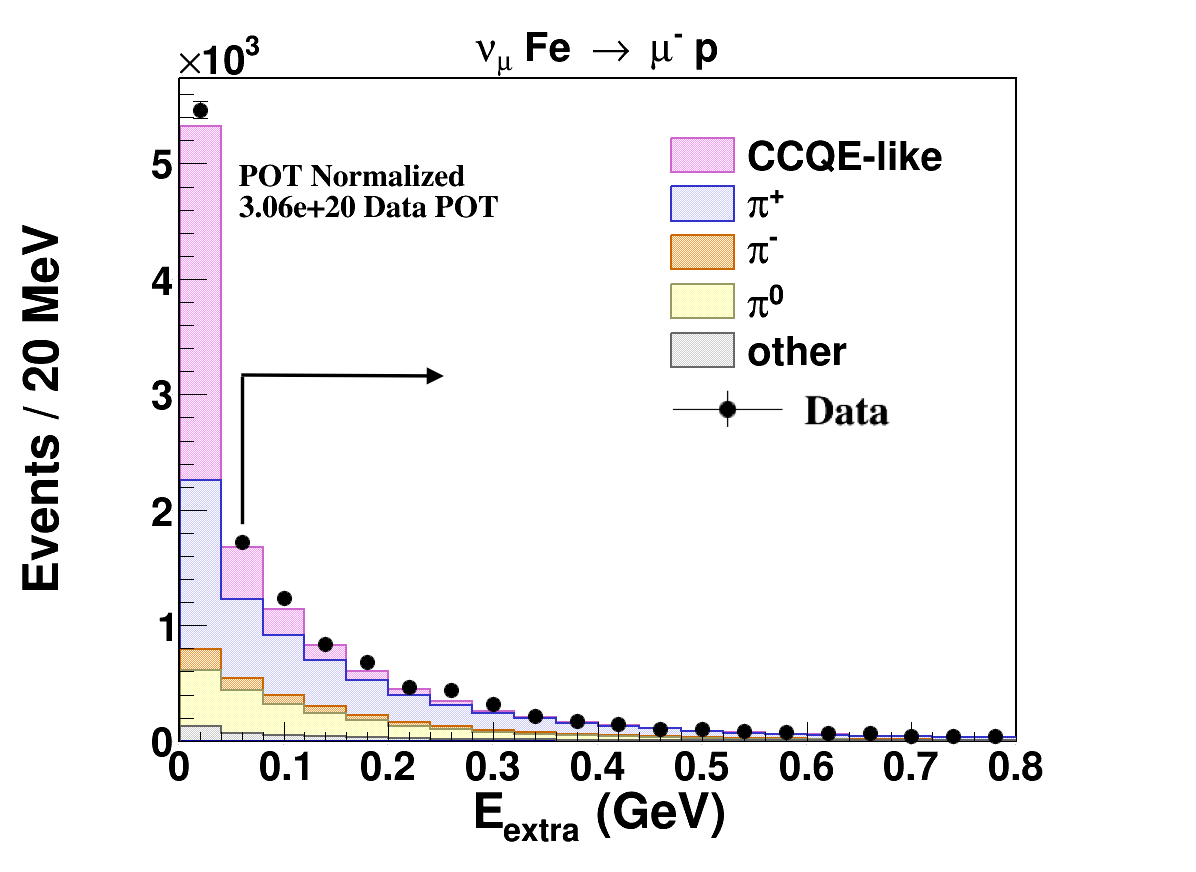}
\includegraphics[trim = 0mm 6mm 15mm 2mm, clip, width=0.46\columnwidth]{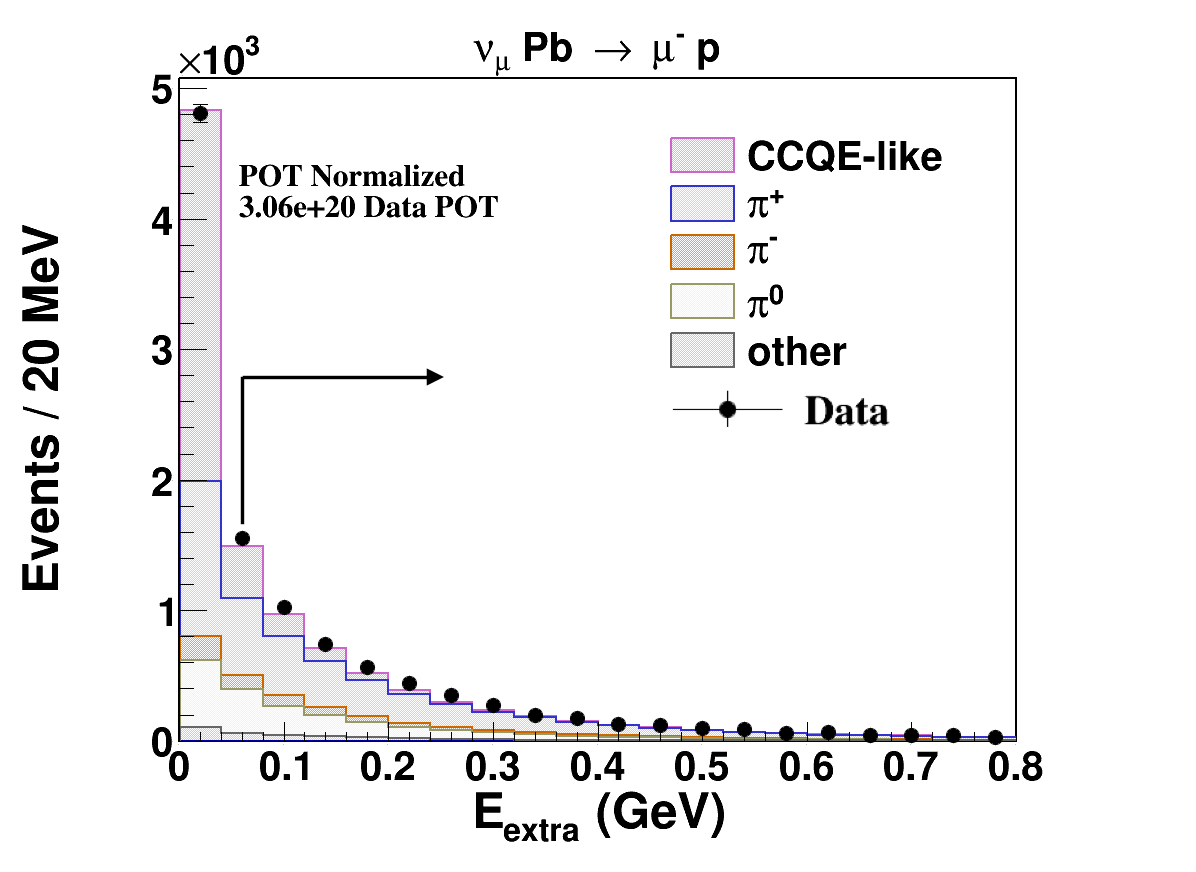}
\caption{Data and Monte Carlo comparison of the $E_{extra}$ distribution for events in Fe (left) and Pb (right). The MC is after background tuning and is normalized to the same number of protons on target. Predictions for CCQE-like, $\pi^+, \pi^-$ and $\pi^0$ in the final state are shown.   }
\label{sideband}\end{figure}

The second background is from interactions that are not QE-like, mostly baryon resonance production, where the pion is misidentified as a proton. This background is constrained by fitting the distribution of events with $E_{extra}>0.05$ GeV shown in Fig. \ref{sideband}. The fit is performed in bins $\QSqproton<0.5$ GeV$^2$ and $\QSqproton>0.5$ GeV$^2$ of four-momentum transfer for each nucleus (C, Fe and Pb) separately. The fit varies the background normalization while keeping the signal constant, until the simulated distributions match the data distributions. Background scale factors span the range from $(1.01\pm{0.10})$ for C to $(1.3\pm{0.1})$ for Pb. As will be elaborated, pion FSI according to some models introduces a significant $A$-dependence for the QE-like signal, and these same FSI could play a similar role in the background constraint samples.

\begin{figure}[htpb]
\centering
\includegraphics[trim = 0mm 4mm 10mm 2mm, clip, width=0.46\columnwidth]{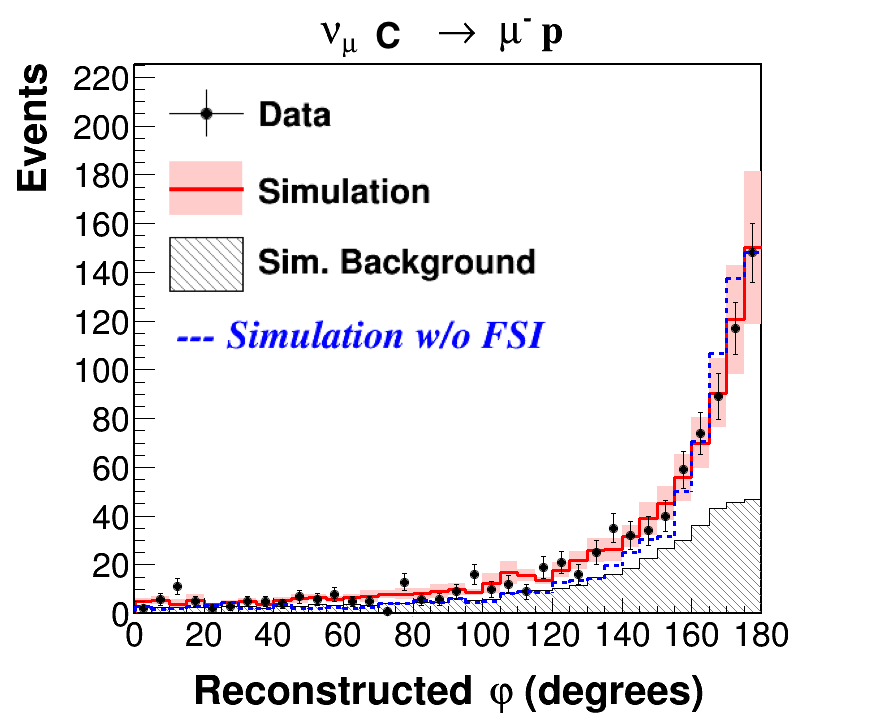}
\includegraphics[trim = 0mm 4mm 10mm 2mm, clip, width=0.46\columnwidth]{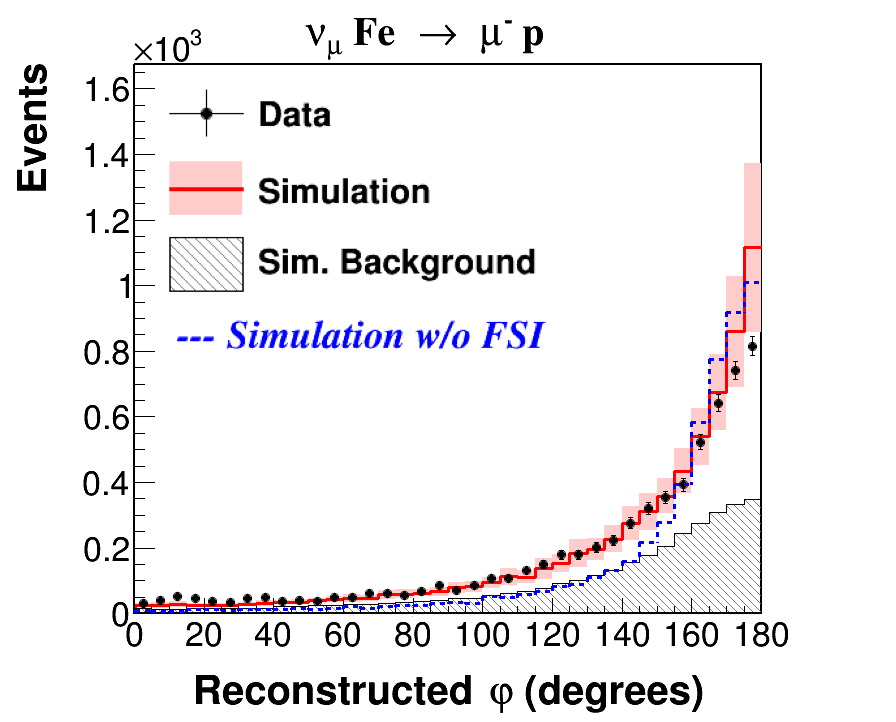}
\includegraphics[trim = 0mm 4mm 10mm 2mm, clip, width=0.46\columnwidth]{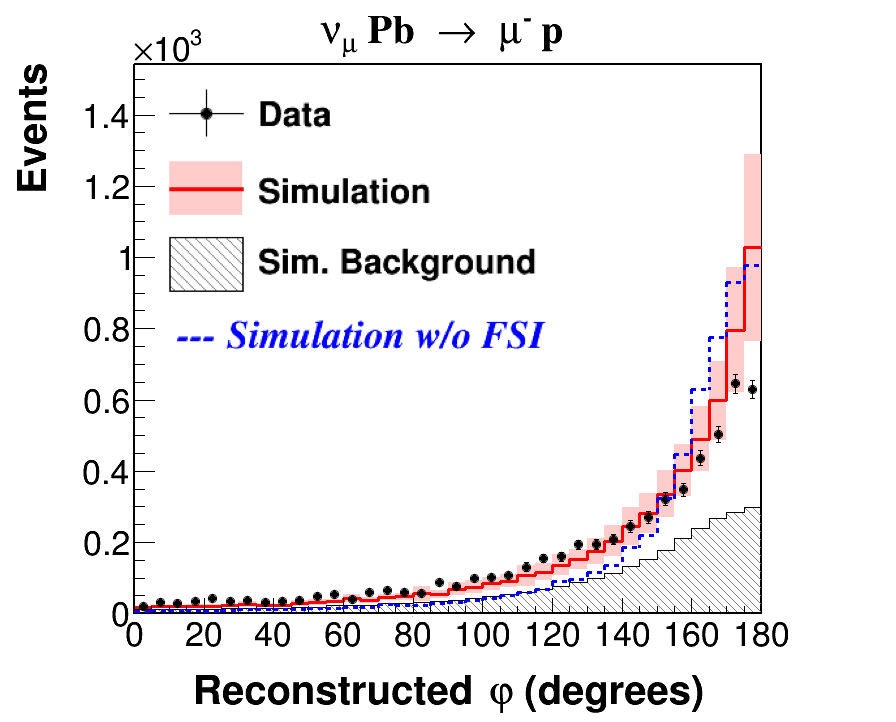}
\includegraphics[trim = 0mm 4mm 10mm 2mm, clip, width=0.46\columnwidth]{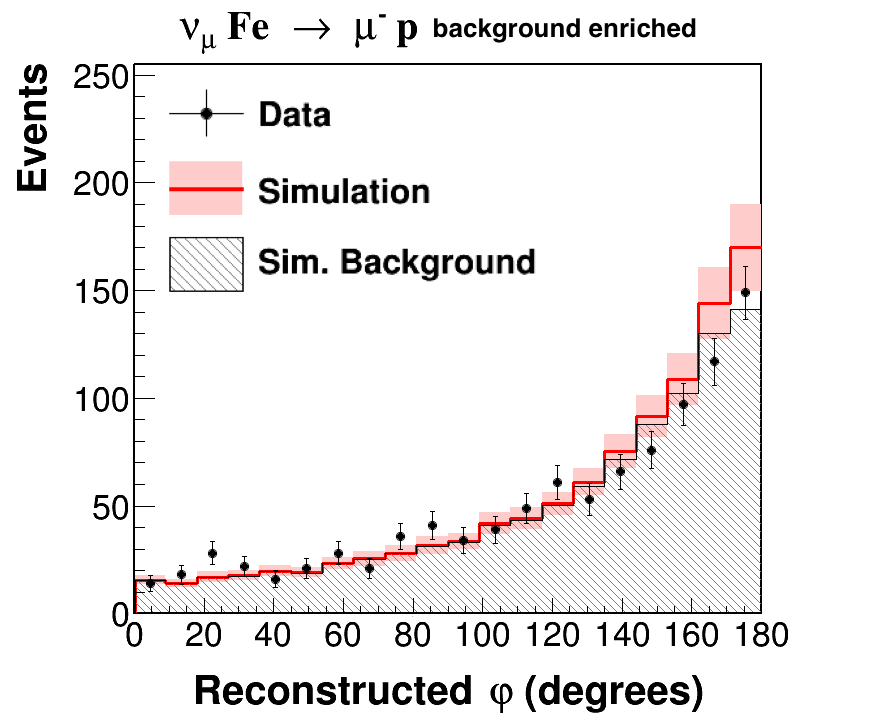}
\caption{ Angle between the neutrino-muon and neutrino-proton planes for the selected events in each nucleus; data and simulation for interactions on carbon (top-left), iron (top-right) and lead (bottom-left). The MC is after background tuning and is normalized to the same number of protons on target. The bottom right plot shows the same distribution in background events, which are dominated by events with a $\pi^+$.}
\label{fig:coplanarity_POT}
\end{figure}

Separating FSI from initial state nuclear effects is challenging since only the combined effects of both are actually measured.  
The coplanarity $\varphi$, the angle between the $\nu$-$\mu$ and $\nu$-p reaction planes, is sensitive to FSI. For a two-body interaction with a neutron at rest, $\varphi=180^{\circ}$. The detector resolution in $\varphi$ is 3.8$^{\circ}$.  Figure \ref{fig:coplanarity_POT} shows $\varphi$ for events passing the CCQE-like selection for  C (top left), Fe (top right), and Pb (bottom left). The simulation predicts that $30\%$ ($10\%$) of the data is backgrounds from resonance production (deep inelastic scattering); some signal also comes from those processes, plus CCQE-like and 2p2h reactions. The width of the distribution is due to Fermi motion, inelastic scattering, and FSI. FSI broadens the distribution by changing the proton direction and by adding a non-QE component. The distribution without FSI is shown to demonstrate the effect FSI has; the no FSI distribution is too narrow and predicts too few events away from the peak at 180 degrees.  The GENIE FSI model appears to describe the broadening for all three samples, but $A$-dependent discrepancies remain for $\varphi$ near 180 degrees.  
To check the modeling of the $\varphi$ distribution for background events, a set of samples of events with a reconstructed Michel electron has been examined, and good agreement is found in each. One example, for Fe, is shown in the bottom right of Fig.~\ref{fig:coplanarity_POT} with the tuned background scale applied.

The differential cross section in \QSqproton bin $i$ is calculated using $(\frac{d\sigma}{dQ^2_p})_i=\frac{\sum{U}_{ij}(N_j-B_j)}{\epsilon_i{T}\Phi\Delta(Q^2_p)_i}$, where $N_j$ is the number selected events, $B_j$ is the estimated number of total background events, $\epsilon_i$ is the signal detector efficiency times acceptance, $T$ is the number of target nuclei, $\Phi$ is the integrated neutrino flux, $\Delta(Q^2_p)$ is the width of bin $i$, and $U_{ij}$ is an operation that accounts for the detector smearing of reconstructed \QSqproton. An unfolding method from~\cite{D'Agostini:1994zf} is used; four iterations are performed to obtain the map from reconstructed to true \QSqproton.

%\section{Systematic Uncertainties} 

\begin{figure}[htpb]
\centering
\includegraphics[trim = 0mm 4mm 10mm 2mm, clip, width=0.48\columnwidth]{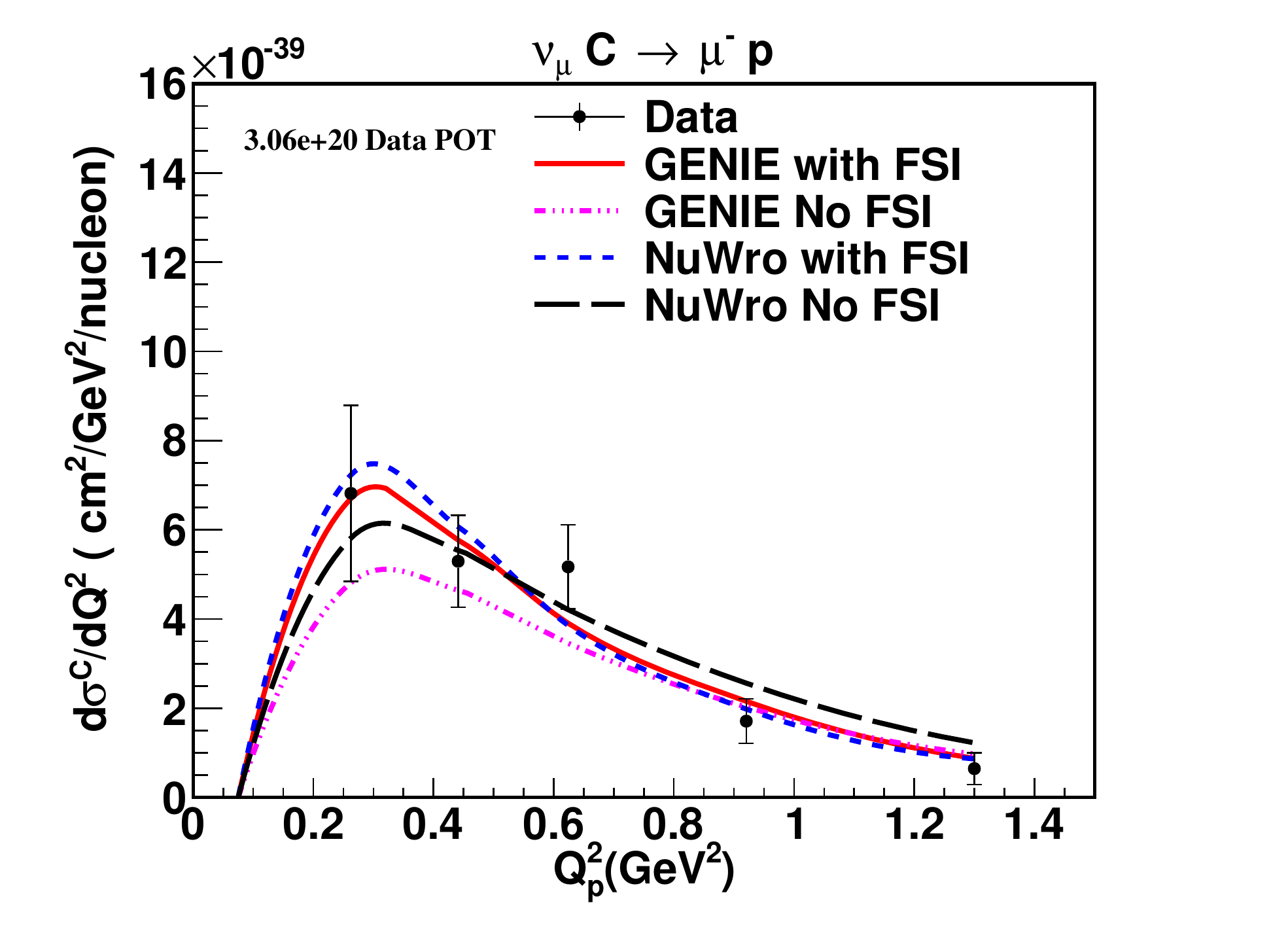}
\includegraphics[trim = 0mm 4mm 10mm 2mm, clip, width=0.48\columnwidth]{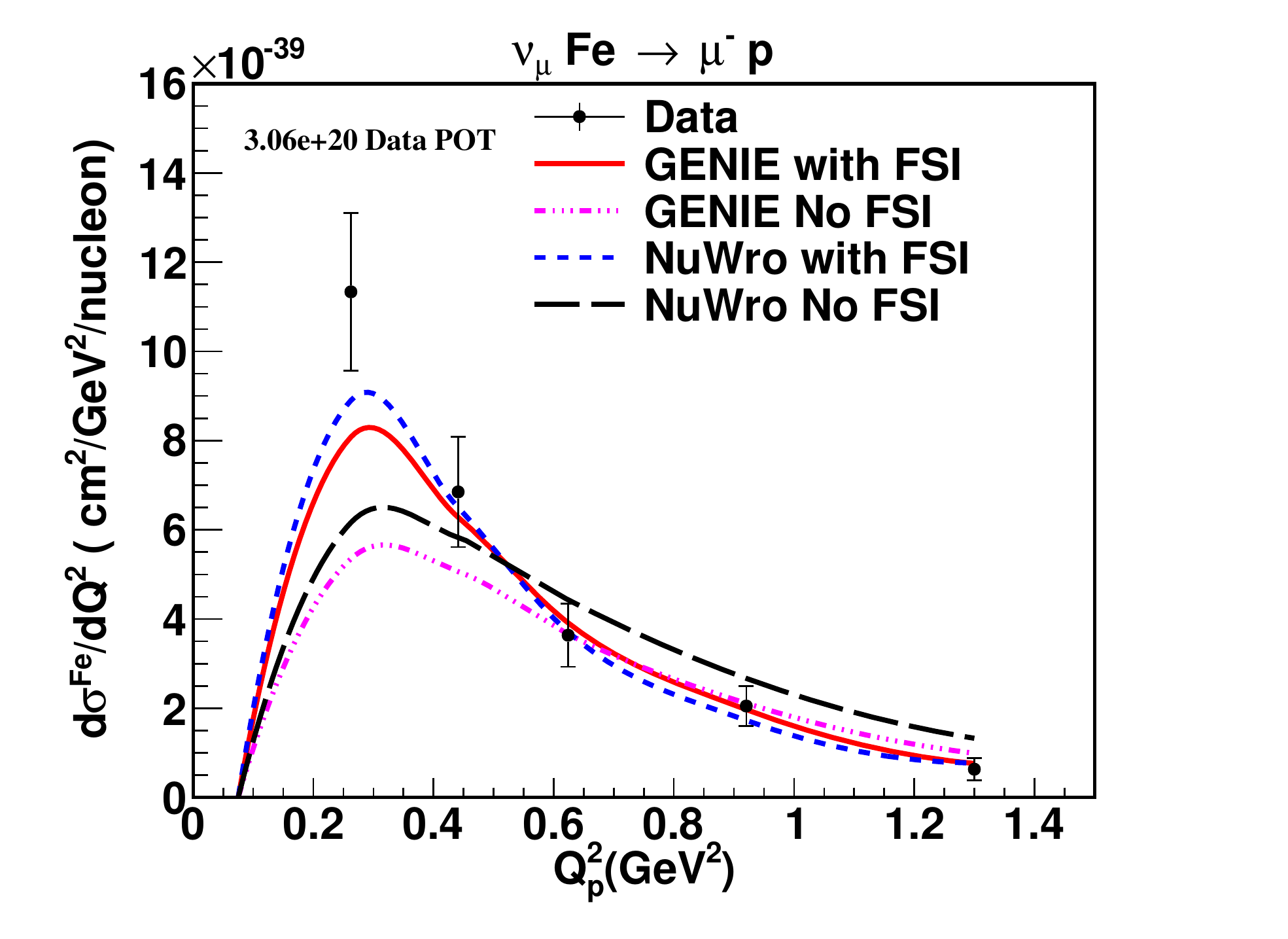}
\includegraphics[trim = 0mm 4mm 10mm 2mm, clip, width=0.48\columnwidth]{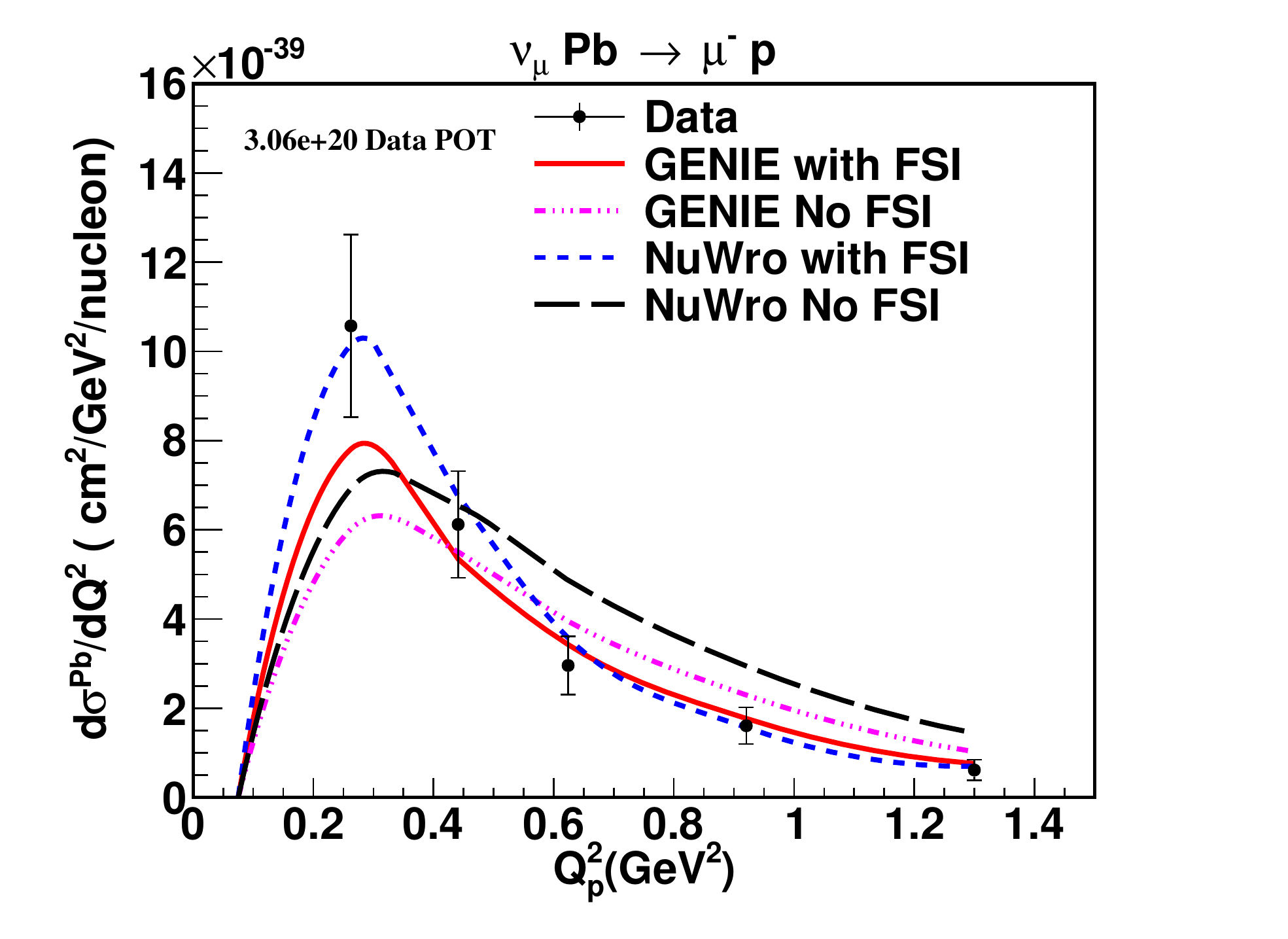}
\includegraphics[trim = 0mm 4mm 10mm 2mm, width=0.48\columnwidth]{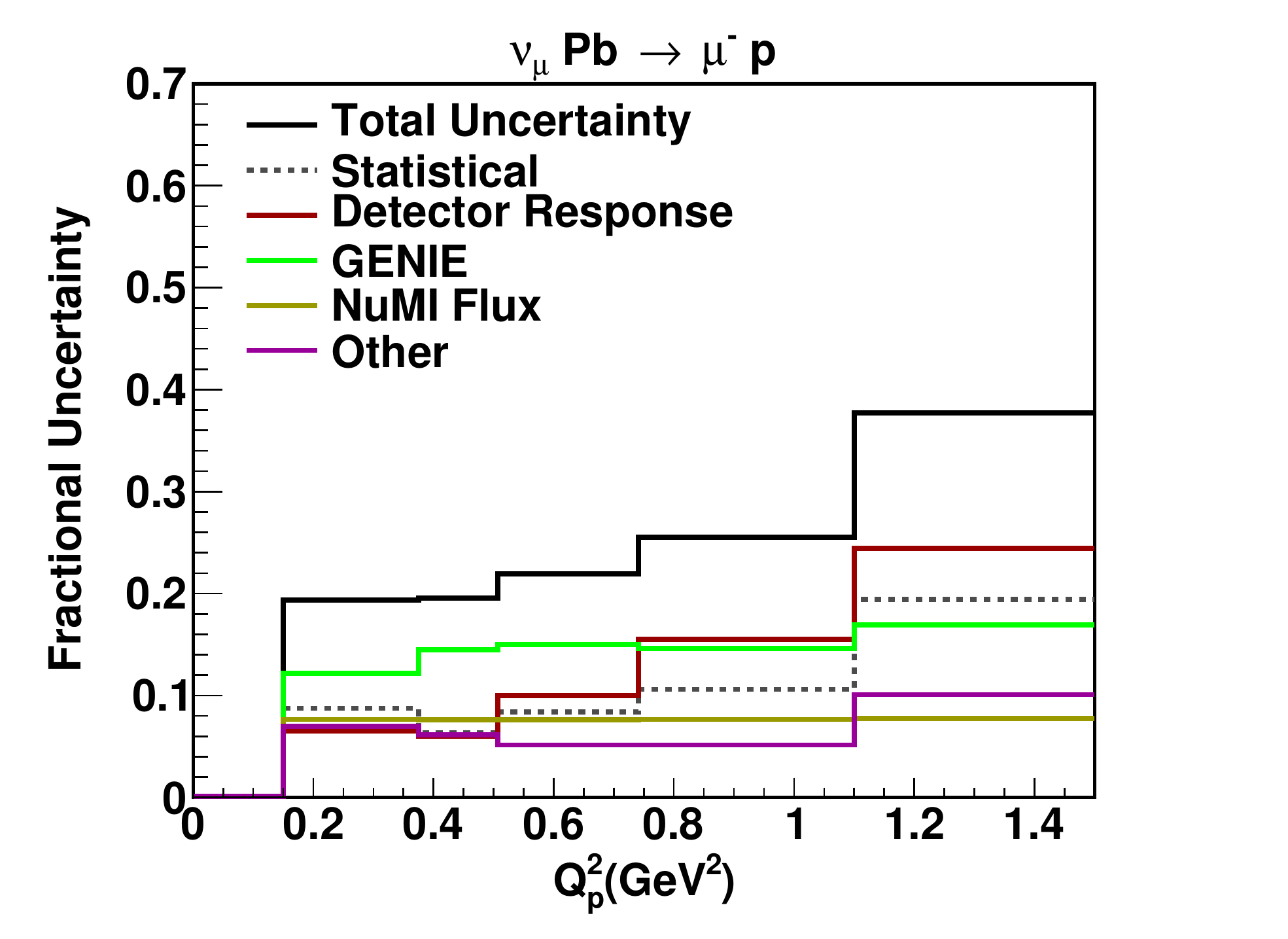}
\caption{ Differential cross sections as a function of $Q^{2}$ for C (top-left), Fe (top-right) and Pb (bottom-left) compared to predictions from \genie and \nuwro which include 2p2h and RPA. The bottom right plot shows the fractional uncertainties for $\frac{d\sigma}{dQ^{2}_{p}}$ of Pb; the dashed curve is from statistical and the solid curves show systematic uncertainties for each of the contributions.}\label{fig:cross_section_model}
\end{figure}

Systematic uncertainties on \xsecproton have been assessed for detector energy response (including hadron propagation), GENIE modeling (including FSI modeling and modeling of 2p2h effects) and the NuMI flux which is obtained from comparisons of the hadron production model with data from hadron production experiment NA49~\cite{Aliaga:2016oaz}.
% Other sources of uncertainty include an extra $10\%$ on the fraction of DIS and resonant to account for the difference in background predictions from the background constraint techniques,
A second background constraint technique was used in which the DIS and resonant background components are floated in the fit. This result differs from the main background constraint by $10\%$, which we take as a systematic uncertainty. Also, we assess the uncertainties from the efficiency of the Michel electron cut, and the number of nucleons in the nuclear targets.
  Most uncertainties are evaluated by randomly varying the associated parameters
in the simulation within uncertainties and re-extracting
\xsecproton. 
Uncertainties in the beam flux affect the normalization of \xsecproton
and are correlated across \QSqproton bins. The uncertainties on the
signal neutrino interaction and FSI models affect \xsecproton primarily through the efficiency
correction, and are dominated by uncertainties on the resonance
production axial form factor, pion absorption, and pion inelastic scattering. The uncertainties
associated with hadron propagation in the \minerva detector
are evaluated by reweighting signal and background simulation by 10$\%$, $15\%$, $20\%$ for C, Fe and Pb respectively, based on comparisons between Geant4 and measurements of $\pi,p,n$-nucleus cross-section on nuclei ~\cite{doi:10.1146/annurev.nucl.52.050102.090713,PhysRevC.23.2173,ALLARDYCE19731,PhysRevC.53.1745}. The assigned systematic uncertainties are shown in the bottom right of Fig \ref{fig:cross_section_model}.

%\section{Results}
Figure \ref{fig:cross_section_model} shows the differential cross sections \xsecproton as a function of \QSqproton for C, Fe, and Pb in data. The GENIE FSI and the NuWro FSI models (with 2p2h and RPA correction included by both generators) are compared to the data.  Predictions from each of these generators without FSI are also compared to the data. As shown previously by Fig. 3, the FSI treatments are needed to achieve agreement. The data exhibits an $A$-dependence and better agreement with NuWro than GENIE; at higher $A$ the protons move to lower energy, suggesting an increase in  proton energy loss in the nucleus. The measurement tests the FSI treatments,  especially with respect to pion absorption and proton inelastic scattering. NuWro has medium effects for pion absorption FSI that give a strong dependence on $A$, effects that are not included in GENIE.

The consequences of incorporating 2p2h events and RPA kinematic distortion into the GENIE simulation were evaluated. Adding the 2p2h reaction to the default GENIE model changes the predicted cross section by 20\% for C, 21\% for Fe, and 22\% for Pb. Adding the RPA produces a $0.6\%$ change because the most affected events are below the proton tracking threshold of this analysis. Neither the 2p2h nor the RPA model predicts significant $A$ dependence.
The chi-square between the GENIE simulation, with and without 2p2h and RPA, and the data is shown in Table \ref{tab:model_compare}, together with the chi-square for NuWro with 2p2h plus RPA compared to the data.

\begin{figure}[htpb]
\centering
\includegraphics[trim = 0mm 4mm 10mm 2mm, clip, width=0.48\columnwidth]{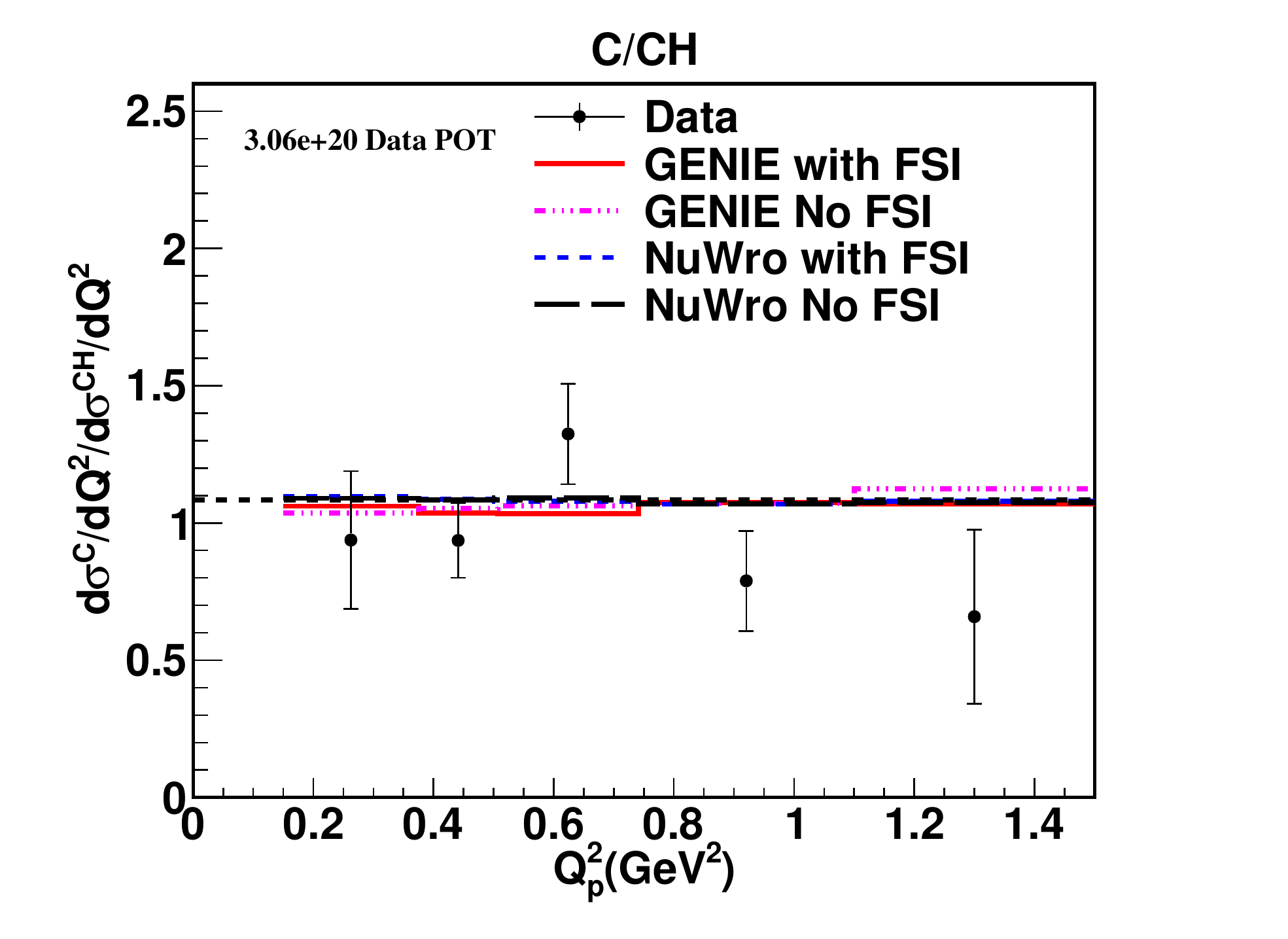}
\includegraphics[trim = 0mm 4mm 10mm 2mm, clip, width=0.48\columnwidth]{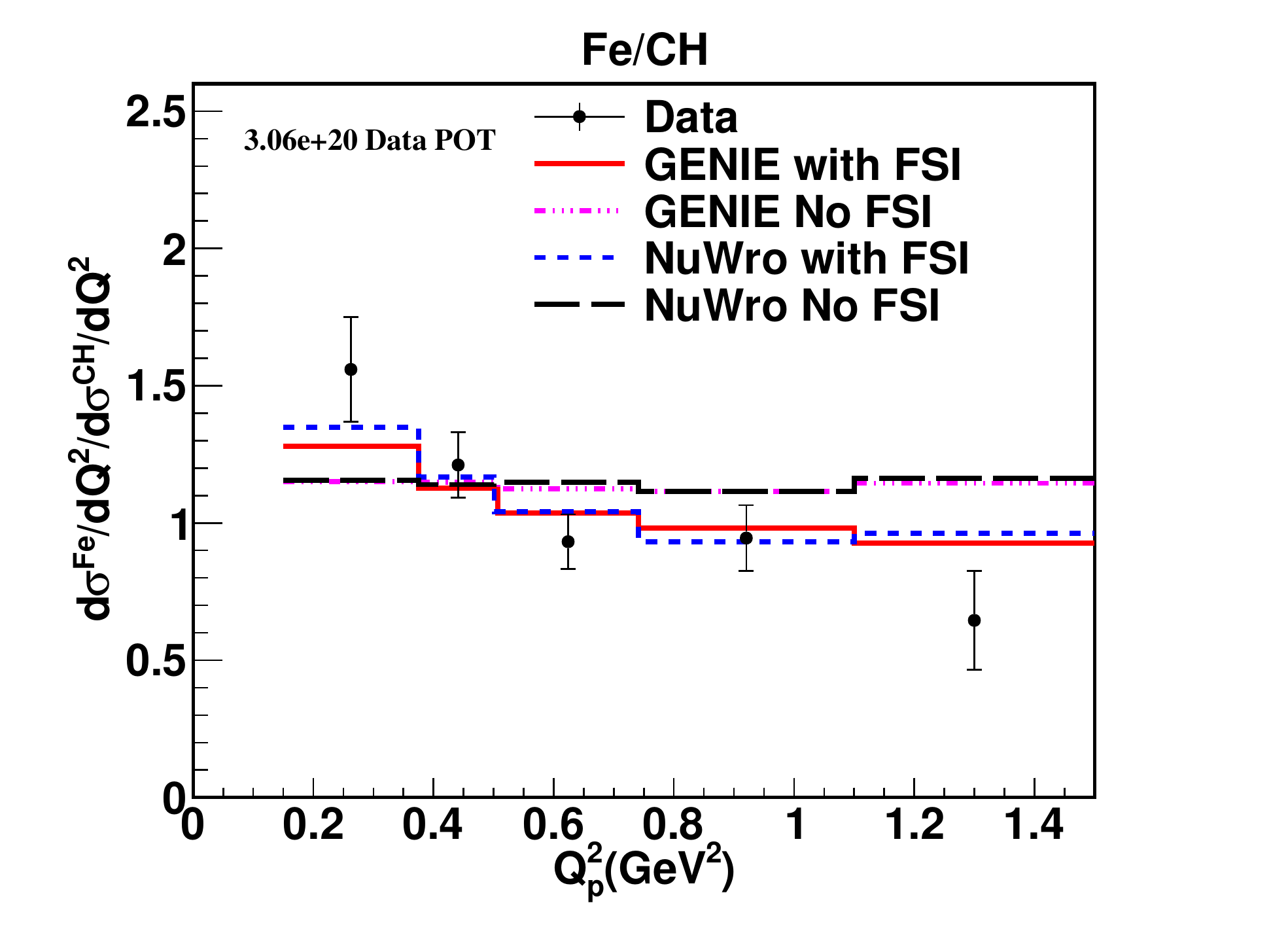}
\includegraphics[trim = 0mm 4mm 10mm 2mm, clip, width=0.48\columnwidth]{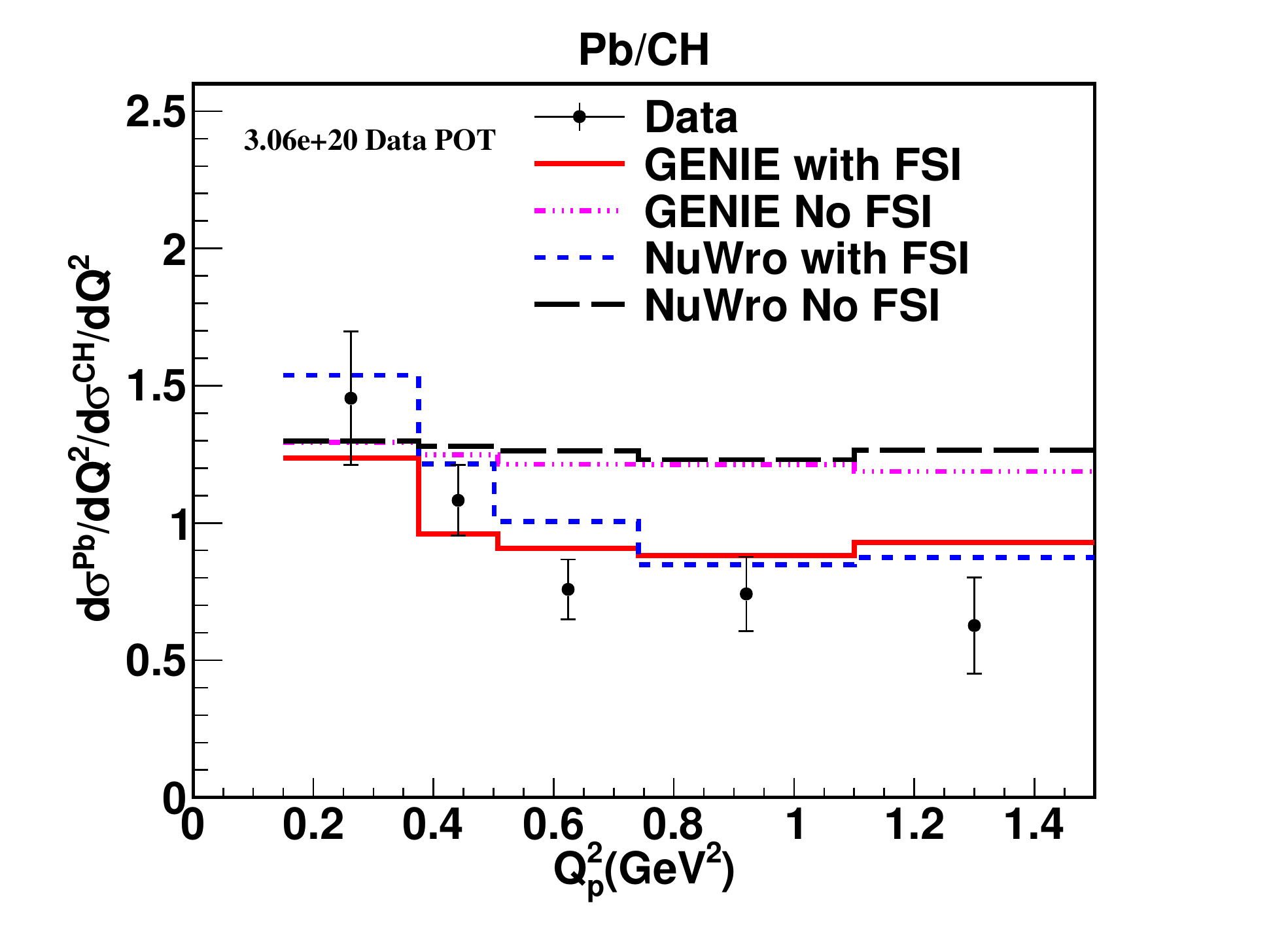}
\includegraphics[trim = 0mm 4mm 10mm 2mm, width=0.48\columnwidth]{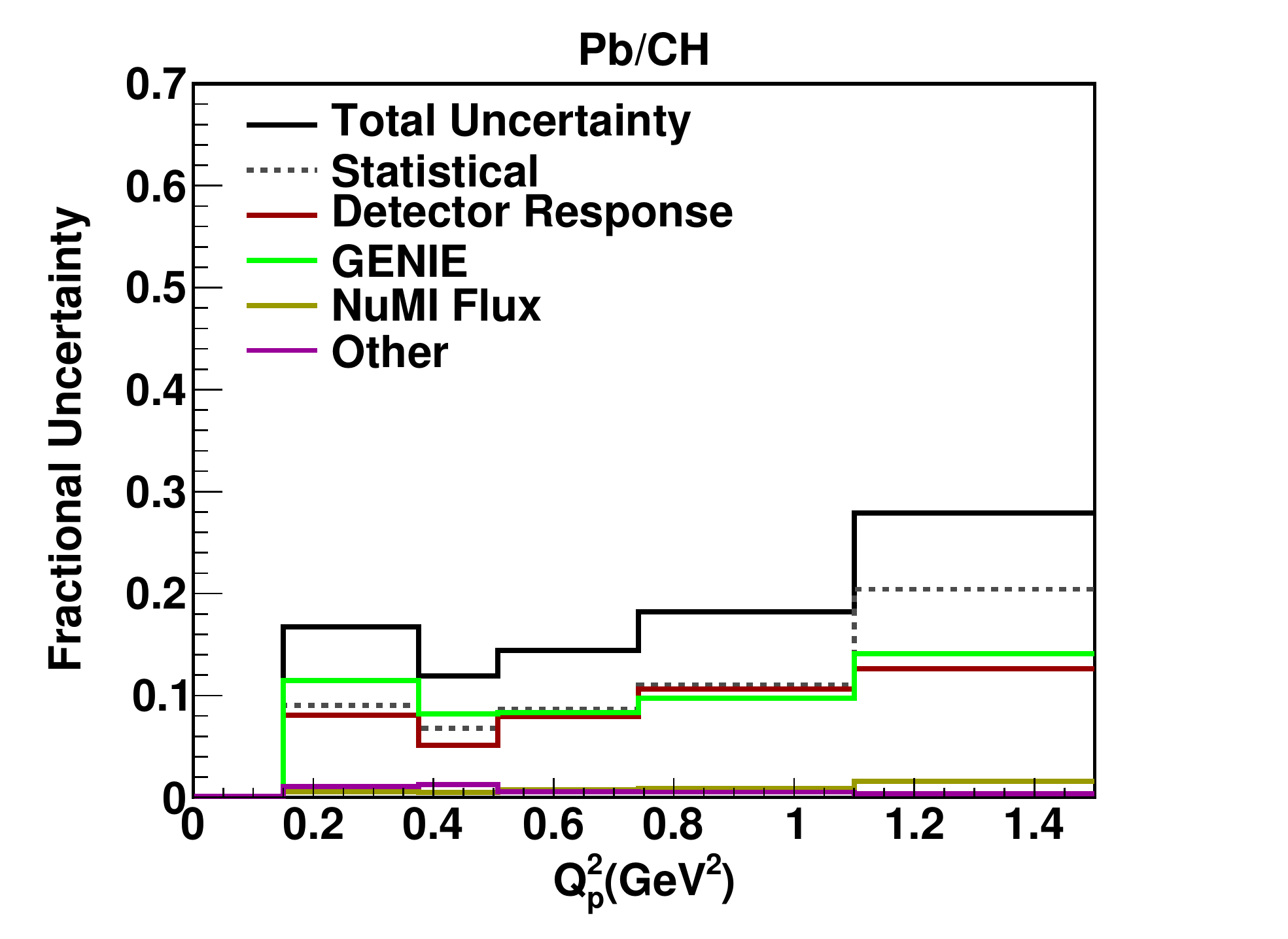}
\caption{ Ratio of $Q^{2}_{p}$ differential cross sections on C(top-left), Fe (top-right) and Pb(bottom-left) to CH, compared to predictions from \genie and \nuwro which include 2p2h and RPA. The bottom right shows the fractional uncertainties for the $\frac{d\sigma}{dQ^{2}_{p}}$ ratio (Pb to CH), the dashed curve shows statistical uncertainty, the solid curves (color online) show systematic uncertainties each of the contributions for the systematics and black is the total uncertainty.}
\label{fig:ratios}
\end{figure}

\begin{table}[!htbp]
\centering
\caption{Calculated $\chi^{2}$ between the data and various models with M$_{A}$ = 0.99~ GeV.
The number of degrees of freedom is 5.\label{tab:model_compare}}
\scalebox{0.8}
{
\begin{tabular}{ l c c c }
               Model & Carbon   & Iron & Lead \\ \hline
\genie  RFG              & 11.0 & 63.8 & 41.1   \\
\genie  RFG + 2p2h       & 5.9  & 18.9 & 16.3   \\
\genie RFG + 2p2h + RPA  & 5.9  & 19.9 & 17.5   \\
\nuwro RFG + 2p2h + RPA  & 6.0  & 14.6 & 11.0   \\
\end{tabular}
}
\end{table}
Figure \ref{fig:ratios} shows the ratios of $\frac{d\sigma}{dQ^{2}_{p}}$ on C, Fe and Pb to the same quantity as measured in the high statistics CH sample.  The data ratios are helped by reduction of systematics uncertainties including the flux. The data ratios emphasize the increasingly strong effect on C, Fe and Pb. The model ratios show that a large effect can be attributed to FSI and is similar for both GENIE and NuWro.  In addition,  NuWro better describes the lowest $Q^2_p$ points with its $A$-dependent pion absorption model and medium corrections. In the $\QSqproton<0.6$ GeV$^2$ region, GENIE predicts that $28\%$ of CCQE-like signal events are from the pion absorption process.
 The coplanarity angle also shows an $A$ dependence, which may partially be from FSI.   

In summary: Quasielastic-like scattering is measured for the first time, in the same neutrino beam, on nuclear targets ( C, Fe, and Pb) that span an $A$-range of 195 nucleons.
The measurements of this work reveal an $A$-dependence for the rate of quasielastic-like scattering which is not well described by current models of neutrino scattering within a nuclear medium. They will serve as benchmarks for the continued refinement of neutrino generators as is required to achieve precise delineations of mass hierarchy and CP nonconservation in the neutrino sector.

\ifnum\sizecheck=1
  \newpage
  {\Large Content after here does not count against size of PRL}
  \newpage
\fi
\begin{acknowledgments}
This work was supported by the Fermi National Accelerator Laboratory under US De- partment of Energy contract No. DE-AC02-07CH11359 which included the MINERvA con- struction project. Construction support was also granted by the United States National Science Foundation under Award PHY-0619727 and by the University of Rochester. Sup- port for participating scientists was provided by NSF and DOE (USA), by CAPES and CNPq (Brazil), by CoNaCyT (Mexico), by Proyecto Basal FB 0821, CONICYT PIA ACT1413, Fondecyt 3170845 and 11130133 (Chile), by CONCYTEC, DGI-PUCP and IDI/IGI-UNI (Peru), and by Latin American Center for Physics (CLAF). We thank the MINOS Collaboration for use of its near detector data. We acknowledge the dedicated work of the Fermilab staff responsible for the operation and maintenance of the beamline and detector and the Fermilab Computing Division for support of data processing.

\end{acknowledgments}

\bibliographystyle{apsrev4-1}
\bibliography{ProtonCCQElike}

\ifnum\PRLsupp=1
  \clearpage
  \newcommand{\qsq}{\ensuremath{Q^2_{QE}}\xspace}
\renewcommand{\textfraction}{0.05}
\renewcommand{\topfraction}{0.95}
\renewcommand{\bottomfraction}{0.95}
\renewcommand{\floatpagefraction}{0.95}
\renewcommand{\dblfloatpagefraction}{0.95}
\renewcommand{\dbltopfraction}{0.95}
\setcounter{totalnumber}{5}
\setcounter{bottomnumber}{3}
\setcounter{topnumber}{3}
\setcounter{dbltopnumber}{3}

{\normalsize \appendix{Appendix: Supplementary Material}\hfill\vspace*{4ex}}

\begin{figure}[htpb]
\centering
\includegraphics[trim = 0mm 4mm 10mm 2mm, clip, width=0.48\columnwidth]{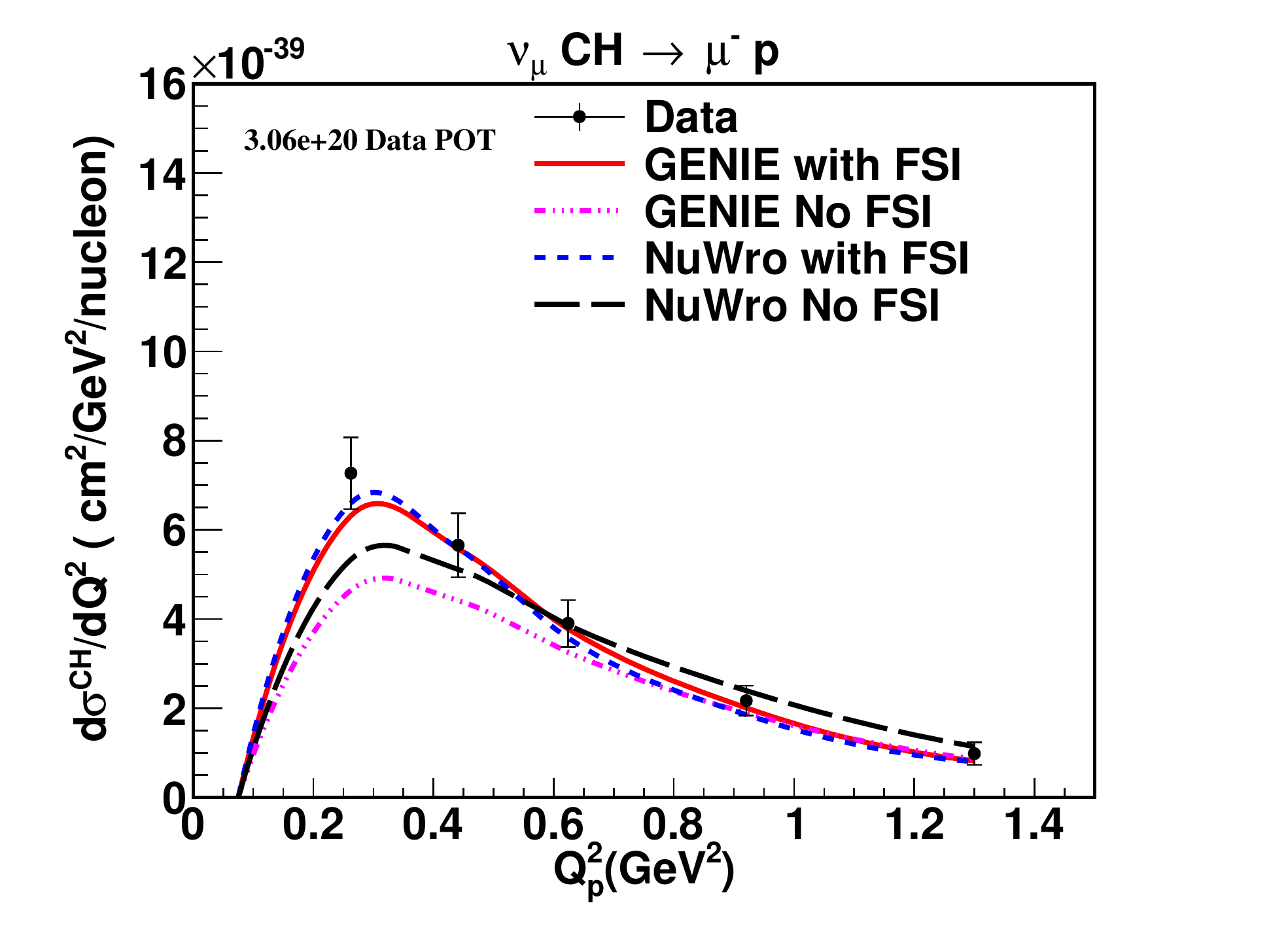}
\includegraphics[trim = 0mm 4mm 10mm 2mm, width=0.48\columnwidth]{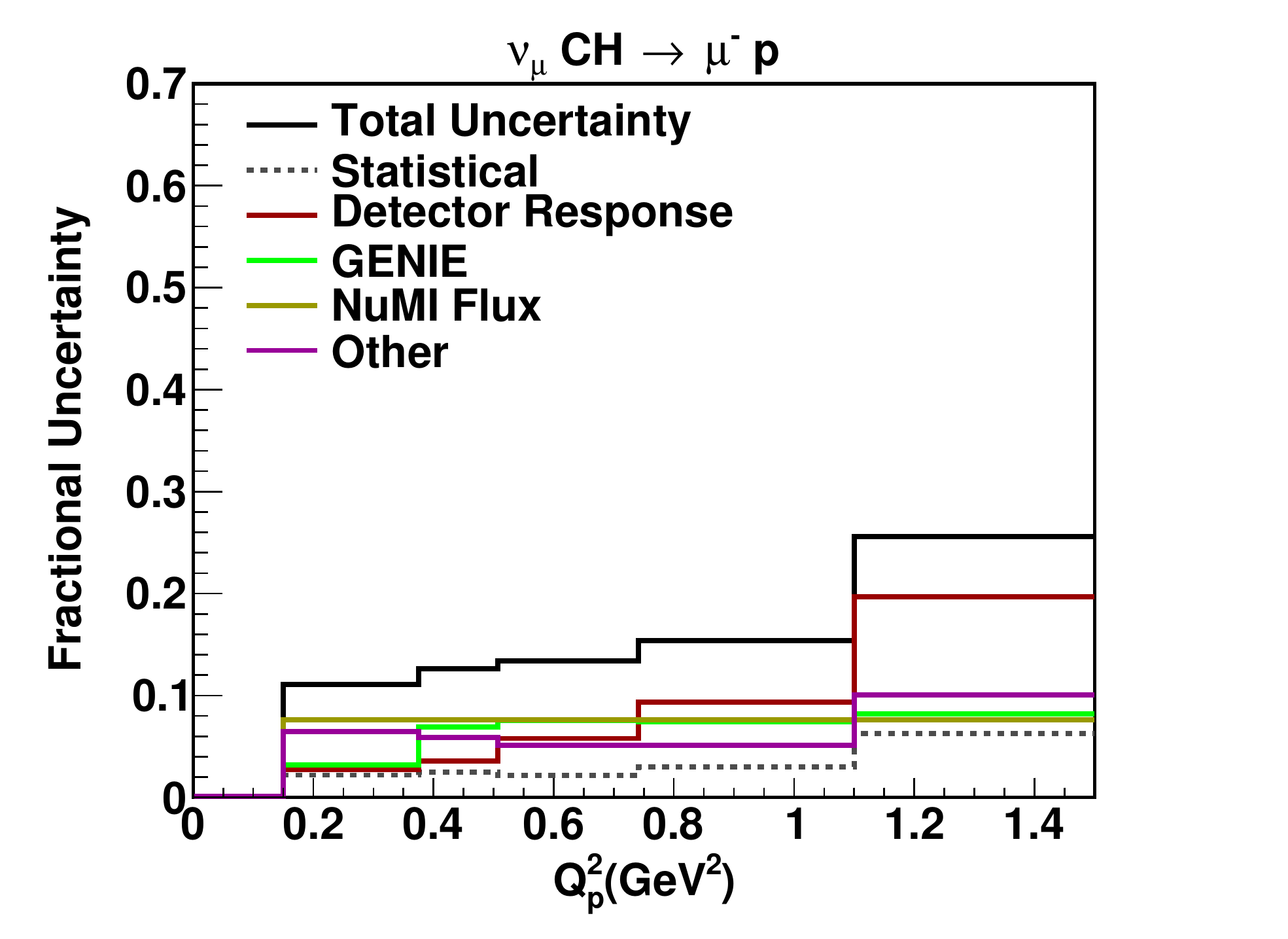}
\caption{Left: Differential cross sections as a function of $Q^{2}_p$ on CH compared to predictions from \genie and \nuwro. Right: Fractional uncertainties for each bin of \QSqproton. The dashed curve shows the statistical uncertainty,  while the solid lines show individual components of the systematic uncertainty.}\label{fig:cross_section_modelCH}
\end{figure}

\begin{figure}[htpb]
\centering
\includegraphics[trim = 2mm 8mm 15mm 8mm, width=0.48\columnwidth]{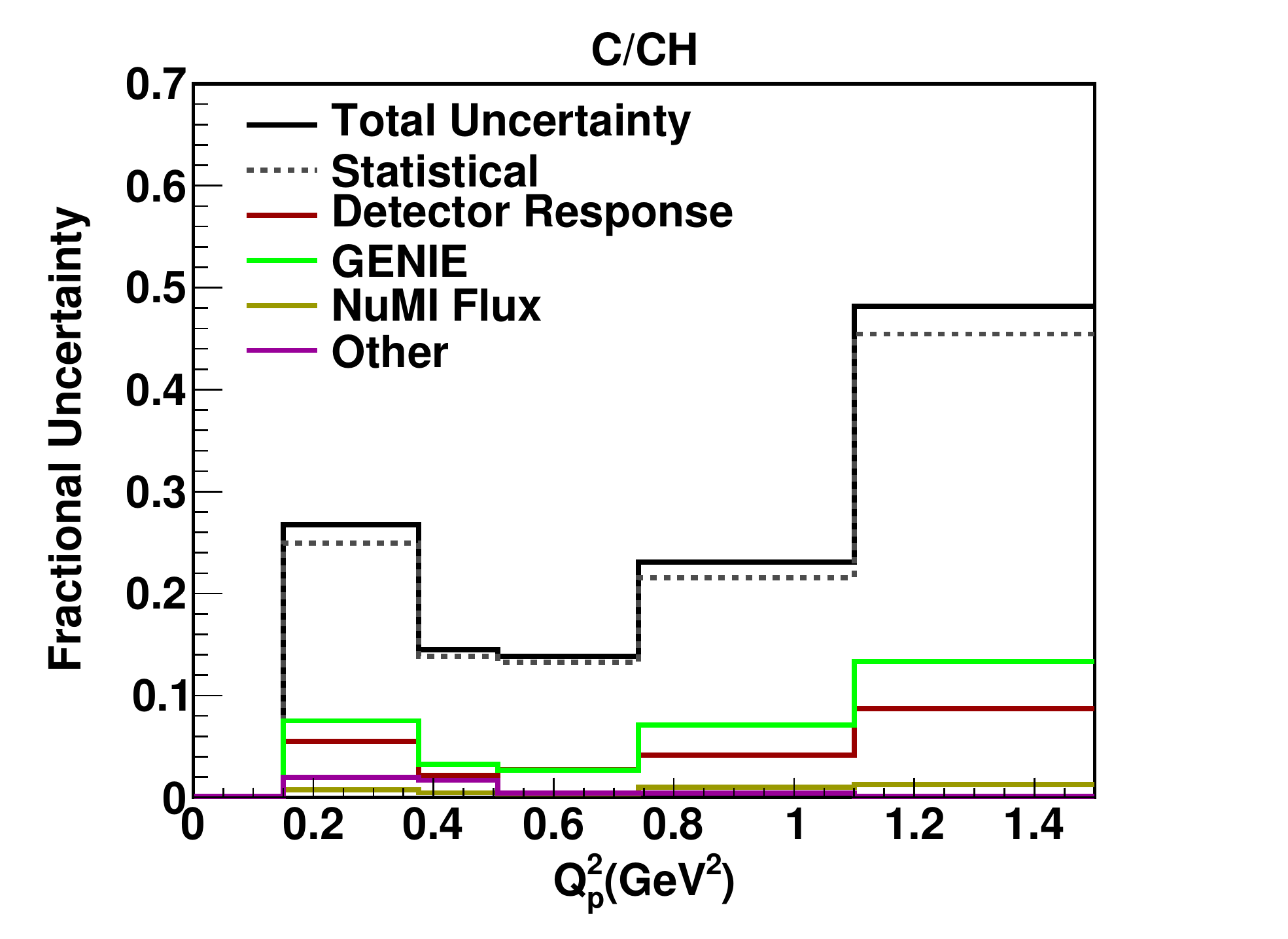}
\includegraphics[trim = 2mm 8mm 15mm 8mm, width=0.48\columnwidth]{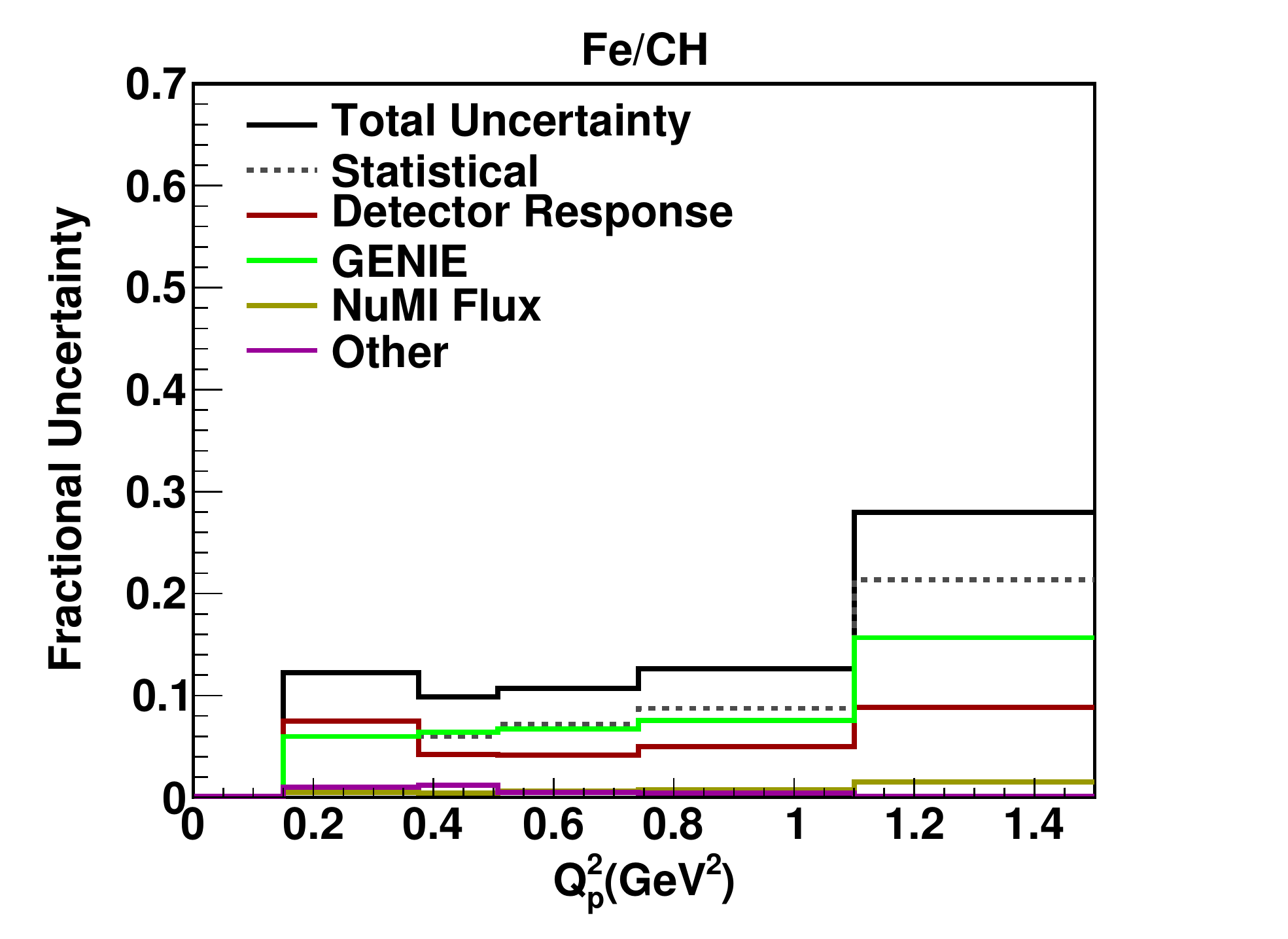}
\caption{ Fractional uncertainties for $\frac{d\sigma}{dQ^{2}_{p}}$ as a function of $Q^2_p$ for C (left) and Fe (right); the dashed curve is from statistical and each of the contributions for the systematics.}\label{fig:cross_section_modelE}
\end{figure}

\begin{figure}[htpb]
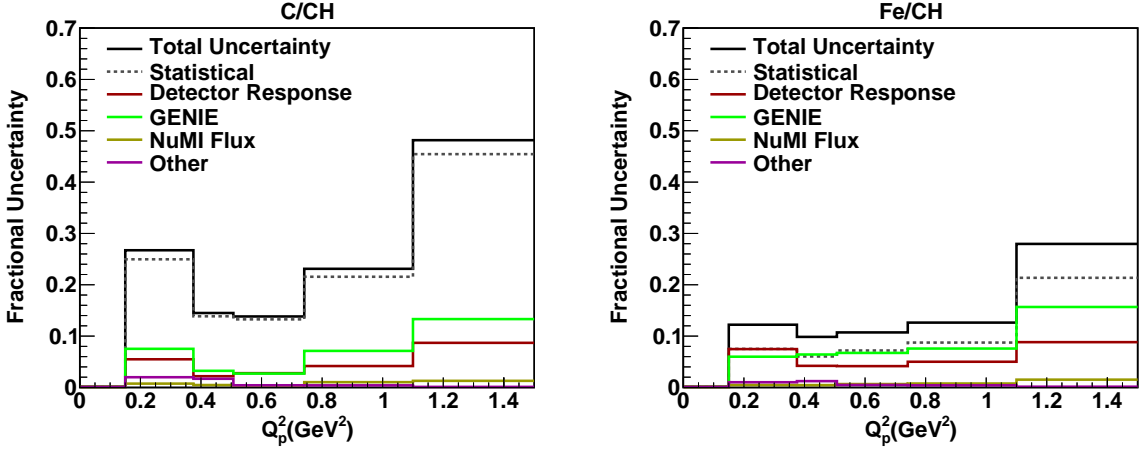

\centering
\includegraphics[trim = 2mm 8mm 15mm 8mm, width=0.48\columnwidth]{Ratio_CError.pdf}
\includegraphics[trim = 2mm 8mm 15mm 8mm, width=0.48\columnwidth]{Ratio_FeError.pdf}
\caption{ Fractional uncertainties for the $\frac{d\sigma}{dQ^{2}_{p}}$ ratio as a function of $Q^2_p$ for C/CH (left) and Fe/CH (right); the dashed curve shows statistical uncertainty, the solid curves (color online) show systematic uncertainties each of the contributions for the systematics and black is the total uncertainty.}
\label{fig:ratiosE}
\end{figure}

\begin{table}[ht]
\begin{center}
    \begin{tabular}{c| c}
    \hline
\hline 
    $Q^{2}_{p}$ ($GeV^2$) & Cross section for C($10^{-38}cm^2/GeV^2$/nucleon) \\  
\hline
    0.15 - 0.375 & 0.682 $\pm$ 0.170 $\pm$ 0.101 \\  
    0.375 - 0.507 & 0.530 $\pm$ 0.072 $\pm$ 0.074\\  
    0.507 - 0.741 & 0.517 $\pm$ 0.068 $\pm$ 0.066\\  
    0.741 - 1.1 & 0.171 $\pm$ 0.037 $\pm$ 0.034\\  
    1.1 - 1.5 & 0.065 $\pm$ 0.029 $\pm$ 0.0206\\  
     \hline
\hline
    \end{tabular}
\caption{Measured \xsecproton and total uncertainties for C }

\end{center}
\end{table}

\begin{table}[ht]
\begin{center}
    \begin{tabular}{c|c}
     \hline
\hline
    $Q^{2}_{p}$ ($GeV^2$) & Cross section for Fe($10^{-38}$$cm^2$/$GeV^2$/nucleon) \\  
\hline
    0.15 - 0.375 & 1.133 $\pm$ 0.081 $\pm$ 0.157 \\  
    0.375 - 0.507 & 0.685 $\pm$ 0.038 $\pm$ 0.117\\  
    0.507 - 0.741 & 0.364 $\pm$ 0.025 $\pm$ 0.066\\  
    0.741 - 1.1 & 0.205 $\pm$ 0.017 $\pm$ 0.042\\  
    1.1 - 1.5 & 0.063 $\pm$ 0.013 $\pm$ 0.022\\  
     \hline
\hline
    \end{tabular}
\caption{Measured \xsecproton and total uncertainties for Fe }
\end{center}
\end{table}

\begin{table}[ht]
\begin{center}
    \begin{tabular}{c|c}
     \hline
\hline
    $Q^{2}_{p}$ ($GeV^2$) & Cross section for Pb ($10^{-38}$$cm^2$/$GeV^2$/nucleon) \\  
\hline
    0.15 - 0.375 & 1.057 $\pm$ 0.092 $\pm$ 0.183 \\  
    0.375 - 0.507 & 0.612 $\pm$ 0.039 $\pm$ 0.113\\  
    0.507 - 0.741 & 0.296 $\pm$ 0.025 $\pm$ 0.060\\  
    0.741 - 1.1 & 0.161 $\pm$ 0.017 $\pm$ 0.037\\  
    1.1 - 1.5 & 0.061 $\pm$ 0.012 $\pm$ 0.020\\  
     \hline
\hline
    \end{tabular}
\caption{Measured \xsecproton and total uncertainties for Pb }
\end{center}
\end{table}

\begin{table}[ht]
\begin{center}
    \begin{tabular}{c|c}
     \hline
\hline
    $Q^{2}_{p}$ ($GeV^2$) & Ratio C/CH \\  
\hline
    0.15 - 0.375 & 0.938 $\pm$ 0.234 $\pm$ 0.089 \\  
    0.375 - 0.507 & 0.94 $\pm$ 0.13 $\pm$ 0.04\\  
    0.507 - 0.741 & 1.32 $\pm$ 0.18 $\pm$ 0.0518\\  
    0.741 - 1.1 & 0.789 $\pm$ 0.170 $\pm$ 0.066\\  
    1.1 - 1.5 & 0.659 $\pm$ 0.299 $\pm$ 0.105\\  
     \hline
\hline
    \end{tabular}
\caption{Measured Ratio C/CH and total uncertainties }
\end{center}
\end{table}

\begin{table}[ht]
\begin{center}
    \begin{tabular}{c|c}
     \hline
\hline
    $Q^{2}_{p}$ ($GeV^2$) & Ratio Fe/CH \\  
\hline
    0.15 - 0.375 & 1.559 $\pm$ 0.117 $\pm$ 0.150 \\  
    0.375 - 0.507 & 1.212 $\pm$ 0.073 $\pm$ 0.0943\\  
    0.507 - 0.741 & 0.932 $\pm$ 0.067 $\pm$ 0.074\\  
    0.741 - 1.1 & 0.945 $\pm$ 0.083 $\pm$ 0.086\\  
    1.1 - 1.5 & 0.645 $\pm$ 0.138 $\pm$ 0.116\\  
     \hline
\hline
    \end{tabular}
\caption{Measured Ratio Fe/CH and total uncertainties }
\end{center}
\end{table}

\begin{table}[ht]
\begin{center}
    \begin{tabular}{c|c}
     \hline
\hline
    $Q^{2}_{p}$ ($GeV^2$) & Ratio Pb/CH  \\  
\hline
    0.15 - 0.375 & 1.454 $\pm$ 0.131 $\pm$ 0.205 \\  
    0.375 - 0.507 & 1.082 $\pm$ 0.074 $\pm$ 0.106\\  
    0.507 - 0.741 & 0.758 $\pm$ 0.066 $\pm$ 0.087\\  
    0.741 - 1.1 & 0.742 $\pm$ 0.082 $\pm$ 0.107\\  
    1.1 - 1.5 & 0.626451 $\pm$ 0.127979 $\pm$ 0.119\\  
     \hline
\hline
    \end{tabular}
\caption{Measured Ratio Pb/CH and total uncertainties }
\end{center}
\end{table}

\begin{table}[ht]
\begin{tabular}{c | c c c c c c | }
\hline
\hline
$Q^2_p$ Bins($GeV^2$)&0.15 - 0.375&0.375 - 0.507&0.507 - 0.741&0.741 - 1.1&1.1 - 1.5\\
\hline
0.15 - 0.375 & 1 & 0.78 & 0.66 & 0.57 & 0.40\\ 
0.375 - 0.507 & & 1 & 0.93 & 0.82 & 0.62 \\ 
0.507 - 0.741 & & & 1 & 0.92 & 0.75 \\ 
0.741 - 1.1 & & & &  1 & 0.86 \\ 
1.1 - 1.5 & & & & &  1  \\ 
\hline
\hline
\end{tabular}
\caption{Correlation matrix for CH}
\end{table}

\begin{table}[ht]
\begin{tabular}{c | c c c c c c | }
\hline
\hline
$Q^2_p$ Bins($GeV^2$)&0.15 - 0.375&0.375 - 0.507&0.507 - 0.741&0.741 - 1.1&1.1 - 1.5\\
\hline
0.15 - 0.375 & 1 & 0.20 & 0.21 & 0.14 & 0.09\\
0.375 - 0.507 & & 1 & 0.46 & 0.38 & 0.23 \\
0.507 - 0.741 & & & 1 & 0.41 & 0.27 \\
0.741 - 1.1 & & & &  1 & 0.36 \\
1.1 - 1.5 & & & & &  1  \\
\hline
\hline
\end{tabular}
\caption{Correlation matrix for C}
\end{table}

\begin{table}[ht]
\begin{tabular}{c | c c c c c c | }
\hline
\hline
$Q^2_p$ Bins($GeV^2$)&0.15 - 0.375&0.375 - 0.507&0.507 - 0.741&0.741 - 1.1&1.1 - 1.5\\
\hline
0.15 - 0.375 & 1 & 0.74 & 0.66 & 0.57 & 0.39\\
0.375 - 0.507 & & 1 & 0.85 & 0.75 & 0.57 \\
0.507 - 0.741 & & & 1 & 83 & 0.67 \\
0.741 - 1.1 & & & &  1 & 0.76 \\
1.1 - 1.5 & & & & &  1  \\
\hline
\hline
\end{tabular}
\caption{Correlation matrix for Fe}
\end{table}

\begin{table}[ht]
\begin{tabular}{c | c c c c c c | }
\hline
\hline
$Q^2_p$ Bins($GeV^2$)&0.15 - 0.375&0.375 - 0.507&0.507 - 0.741&0.741 - 1.1&1.1 - 1.5\\
\hline
0.15 - 0.375 & 1 & 0.78 & 0.65 & 0.46 & 0.30\\
0.375 - 0.507 & & 1 & 0.80 & 0.62 & 0.43 \\
0.507 - 0.741 & & & 1 & 0.77 & 0.60 \\
0.741 - 1.1 & & & &  1 & 0.74 \\
1.1 - 1.5 & & & & &  1  \\
\hline
\hline
\end{tabular}
\caption{Correlation matrix for Pb}
\end{table}

\begin{table}[ht]
\begin{tabular}{c | c c c c c c | }
\hline
\hline
$Q^2_p$ Bins($GeV^2$)&0.15 - 0.375&0.375 - 0.507&0.507 - 0.741&0.741 - 1.1&1.1 - 1.5\\
\hline
0.15 - 0.375 & 1 & 0.53 & 0.43 & 0.34 & 0.14\\
0.375 - 0.507 & & 1 & 0.52 & 0.44 & 0.24 \\
0.507 - 0.741 & & & 1 & 0.50 & 0.35 \\
0.741 - 1.1 & & & &  1 & 0.40 \\
1.1 - 1.5 & & & & &  1  \\
\hline
\hline
\end{tabular}
\caption{Correlation matrix for Ratio Fe/CH  }
\end{table}
\newpage
\begin{table}[ht]
\begin{tabular}{c | c c c c c c|  }
\hline
\hline
$Q^2_p$ Bins($GeV^2$)&0.15 - 0.375&0.375 - 0.507&0.507 - 0.741&0.741 - 1.1&1.1 - 1.5\\
\hline
0.15 - 0.375 & 1 & 0.61 & 0.43 & 0.22 & 0.03\\
0.375 - 0.507 & & 1 & 0.54 & 0.35 & 0.14 \\
0.507 - 0.741 & & & 1 & 0.57 & 0.38 \\
0.741 - 1.1 & & & &  1 & 0.48 \\
1.1 - 1.5 & & & & &  1  \\
\hline
\hline
\end{tabular}
\caption{Correlation matrix for Ratio Pb/CH  }
\end{table}

%\begin{table}[ht]
%\begin{tabular}{c | c c c c c c  }
%\hline
%\hline
%$Q^2_p$ Bins($GeV^2$)&0.15 - 0.375&0.375 - 0.507&0.507 - 0.741&0.741 - 1.1&1.1 - 1.5\\
%\hline
%0.15 - 0.375 & 1 & 0.61 & 0.43 & 0.22 & 0.03\\
%0.375 - 0.507 & & 1 & 0.54 & 0.35 & 0.14 \\
%0.507 - 0.741 & & & 1 & 0.57 & 0.38 \\
%0.741 - 1.1 & & & &  1 & 0.48 \\
%1.1 - 1.5 & & & & &  1  \\
%\hline
%\hline
%\end{tabular}
%\caption{Ratio Carbon/CH}
%\end{table}

\fi

\end{document}